\documentclass[aps,prd,preprint,groupedaddress,11pt,nofootinbib]{revtex4-2}  %% for pdflatex  (without dvipdfmx)

\usepackage{hyperref,graphics}
\usepackage{amsmath,amssymb,braket}
\usepackage{mathtools}
\usepackage{fancyvrb,comment,array}
\usepackage{booktabs,multirow}

\allowdisplaybreaks[1]
\usepackage{color}
\newcommand{\Tr}{\mathrm{Tr}}
\newcommand{\diag}{\mathrm{diag}}

\newcommand{\LC}{\mathrm{LC}}
\newcommand{\HM}{\mathrm{HM}}
\renewcommand{\L}{\mathcal{L}}
\renewcommand{\O}{\mathcal{O}}
\newcommand{\A}{\mathcal{A}}

\newcommand\vecsta[1]{{\vec{#1}}^{\,\,{*}}}
\newcommand\vecsqu[1]{{\vec{#1}}^{\,\,2}}

\newcommand{\Slash}[1]{\ooalign{\hfil/\hfil\crcr$#1$}}

\def\simge{\mathrel{
\rlap{\raise~0.511ex \hbox{$>$}}{\lower~0.511ex \hbox{$\sim$}}}}
\def\simle{\mathrel{
\rlap{\raise~0.511ex \hbox{$<$}}{\lower~0.511ex \hbox{$\sim$}}}}
\def\bigs{\mathrel{
\rlap{\raise~0.531ex \hbox{$>$}}{\lower~0.531ex \hbox{$<$}}}}

\usepackage{booktabs}
\usepackage[normalem]{ulem}
\usepackage{multirow}
\begin{document}
\preprint{}

\title{Analysis of $T_{cc}$ and $T_{bb}$ based on the hadronic molecular model and their spin multiplets}
\author{Manato~Sakai}
\email[]{msakai@hken.phys.nagoya-u.ac.jp}
\affiliation{Department of Physics, Nagoya University, Nagoya 464-8602, Japan}
\author{Yasuhiro~Yamaguchi}
\email[]{yamaguchi@hken.phys.nagoya-u.ac.jp}
\affiliation{Department of Physics, Nagoya University, Nagoya 464-8602, Japan}
\affiliation{Kobayashi-Maskawa Institute for the Origin of Particles and the Universe, Nagoya University, Nagoya, 464-8602, Japan}
\affiliation{Meson Science Laboratory, Cluster for Pioneering Research, RIKEN, Hirosawa, Wako, Saitama 351-0198, Japan}

\begin{abstract}
  ${T_{cc}(cc\bar{u}\bar{d})}^{+}$ has been reported by the LHCb experiment in 2022. 
The analysis using the Breit-Wigner parametrization found
the small binding energy, $0.273$ MeV, 
which is 
measured from the threshold of $D^{*+}D^{0}$. 
In this paper, we consider $T_{cc}$ as a $DD^*$ hadronic molecule as a deuteron-like state. 
The one boson exchange model is employed as for the heavy meson interactions, where we determine the cut-off parameter $\Lambda$ to reproduce the reported binding energy of $T_{cc}$ with $I(J^P)=0(1^+)$.
 We discuss the properties of the bound state, and also search for
 $T_{cc}$ with the quantum numbers other than $0(1^{+})$.
  Futhermore, we 
 analyze
 $T_{bb}$ as a bottom counterpart of $T_{cc}$, which involves two bottom quarks, and 
 obtain several bound states. 
Finally we consider the light-cloud basis for wave functions of the doubly heavy tetraquarks in the heavy quark limit. Using the basis, we find 
the spin multiplets  
  of 
 their bound states, indicating the spin structures of diquarks in $T_{cc}$ and $T_{bb}$ 
 with the finite quark masses.
\end{abstract}

\maketitle
\section{Introduction}\label{sec;introduction}
Most hadrons can be classified into baryons consisting of three quarks $qqq$ and mesons consisting of a quark and an antiquark $q\bar{q}$. However, hadrons with more than three quarks are not prohibited in the Quantum Chromodynamics (QCD), which was already indicated by Gell-Mann and Zweig~\cite{Gell-Mann:1964ewy,Zweig:1964ruk,Zweig:1964jf}. 
Hadrons like these states which cannot be explained by ordinary hadrons $q\bar{q}$ and $qqq$ are called exotic hadrons. 
In the heavy quark sector, various exotic hadrons such as $X$, $Y$, $Z$ and $P_{c}$ have been reported since the report on $X(3872)$ in 2003~\cite{Brambilla:2010cs,Belle:2003nnu,LHCb:2015yax,LHCb:2019kea,LHCb:2020bwg,LHCb:2021vvq,Chen:2022asf}. The structures and interactions of these states are not well understood and these studies are important subjects of current researches in the hadron physics.\\
\indent
In 2022, ${T_{cc}(cc\bar{u}\bar{d})}^{+}$ has been reported by the LHCb experiment. The Breit-Wigner mass relative to the $D^{*+}D^{0}$ mass threshold $\delta m_{\mathrm{BW}}$ is 
\begin{equation*}
  \delta m_{\mathrm{BW}} = -273\pm 61\pm 5 ^{+11}_{-14}\ \mathrm{keV}/c^2,
\end{equation*}
while the pole mass relative to the $D^{*+}D^{0}$ mass threshold $\delta m_{\mathrm{pole}}$ is 
\begin{equation*}
  \delta m_{\mathrm{pole}} = -360\pm 40^{+4}_{-0}\ \mathrm{keV}/c^{2},
\end{equation*}
respectively~\cite{LHCb:2021vvq,LHCb:2021auc}. 
The charm number is $2$ and the baryon number is $0$, thus this state is considered as a genuine exotic state.
The isospin, spin and parity are considered as $I(J^P)=0(1^+)$. 
$T_{cc}$ has widely been studied before the report on ${T_{cc}(cc\bar{u}\bar{d})}^{+}$ by the LHCb experiment. In 1980's, there were 
theoretical studies by using the nonrelativistic quark model~\cite{Ader:1981db,Ballot:1983iv,Zouzou:1986qh}. 
Many theoretical studies also have been conducted by using many models such as the hadronic molecule~\cite{Ohkoda:2012hv,Li:2012ss,Cheng:2022qcm,Wang:2021yld,Wang:2021ajy,Ren:2021dsi,Asanuma:2023atv}, 
various quark models~\cite{Meng:2020knc,Meng:2021yjr,Yan:2018gik}, 
the heavy quark symmetry~\cite{Eichten:2017ffp,Cheng:2020wxa}, the string model~\cite{Andreev:2021eyj}, the QCD sum rules~\cite{Navarra:2007yw,Du:2012wp,Agaev:2021vur,Xin:2021wcr,Xin:2022bzt},
the Lattice QCD~\cite{Ikeda:2013vwa,Padmanath:2022cvl,Lyu:2023xro} and so on as summarized in recent reviews ~\cite{Chen:2022asf}.

The small binding energy of $T_{cc}$ from the $DD^\ast$ threshold motivates us to study the $DD^\ast$ molecular structure. 
When considering a hadronic molecule with heavy quarks, we respect the heavy quark spin symmetry (HQS)~\cite{Neubert:1993mb,Grinstein:1995uv,Casalbuoni:1996pg}. 
The HQS leads to mass degeneracy of the heavy pseudoscalar and vector mesons, because of suppression of the chromomagnetic interaction in the heavy quark limit.
In fact, the mass difference of $D$ and $D^{*}$ is approximately $140\,\mathrm{MeV}$, 
which is smaller than those in the light quark sectors, e.g. $m_\rho-m_\pi \sim 630\,$MeV and $m_{K^\ast}-m_K \sim 390\,$MeV.
Thus $D$ and $D^{*}$ are considered as the HQS doublet. $B$ and $B^{*}$ are considered in the same way because the mass difference of $B$ and $B^{*}$ is 
approximately $45\,\mathrm{MeV}$. 
Thus, in a hadronic molecular system of two mesons, these facts lead to channel coupling of $PP$, $PP^{*}$ and $P^{*}P^{*}$ ($P=D,B$ and $P^{*}=D^{*},B^{*}$).  \\
\indent
\begin{comment}
Exotic hadrons have been studied by using a diquark model other than a hadronic molecule. The hadronic molecular basis (HMB) can be transformed to the light-cloud basis (LCB) by a unitary transformation~\cite{Yasui:2013vca,Yamaguchi:2014era,Shimizu:2019ptd}. We can see the spin structures of a diquark by the LCB, 
which we cannot by the HMB. We can see the HQS multiplets, indicating the existence of the partners. \\
\end{comment}
The HQS multiplets can be also seen 
in a hadronic molecule by using the light-cloud basis (LCB)~\cite{Yasui:2013vca,Yamaguchi:2014era,Shimizu:2019ptd}. The light-cloud basis can be obtained by a unitary transformation of the hadronic-molecule basis (HMB), where the spin wave function of the hadronic molecule is decomposed into spins of heavy quarks and the light cloud. 
Thus, 
we can see the spin structures of 
quarks in
the LCB, which 
are hidden in the HMB, and the states can be classified by the 
quark spin structures. 
The HQS multiplet of the hadronic molecule with single heavy quark has been discussed
in Ref.~\cite{Yasui:2013vca}. 
In Ref.~\cite{Shimizu:2019ptd}, the author has classified the hidden-charm pentaquark $P_c$ by using the LCB. 
Interestingly the HQS multiplets indicate the existence of partners. Since 
states with the same spin structures 
of diquarks
belong to the same HQS multiplets, 
the existence of one state indicates that of partners.

In this paper, we 
study 
the doubly heavy tetraquarks $T_{cc}$ and the bottom counterpart $T_{bb}$ as a hadronic molecule of two mesons. 
In Ref.~\cite{Ohkoda:2012hv}, the author has studied the doubly heavy tetraquarks with several quantum numbers 
as the hadronic molecules of $D^{(\ast)}D^{(\ast)}$ and $B^{(\ast)}B^{(\ast)}$ before the LHCb observation. This study has used the one pion exchange potential (OPEP) and the one boson exchange potential (OBEP), 
with the addition of $\rho$ and $\omega$. We follow the study in Ref.~\cite{Ohkoda:2012hv}, 
and in addition we also consider the one $\sigma$ meson exchange in the OBEP to 
construct
a more realistic interaction. We determine the cutoff parameter $\Lambda$ for the OBEP to reproduce the empirical binding energy of $T_{cc}$ with $I(J^P)=0(1^+)$. 
We also study the properties of the obtained bound state. 
$T_{cc}$ with quantum numbers other than $0(1^+)$ is also investigated. 
Using the same potentials,
we discuss the existence of $T_{bb}$ as the $B^{(\ast)}B^{(\ast)}$ molecule.
Finally, we apply the LCB to the doubly heavy tetraquarks in the heavy quark limit, where the bound states are classified by their spin structures of heavy diquark and light antidiquark. The obtained HQS multiplet structure indicates the existence of partner states of the tetraquark bound states.

This paper is organized as follows. In Sec.~\ref{sec;formalism}, we introduce the formalisms of the OPEP and the OBEP. In Sec.~\ref{sec;numerical_results}, we show our numerical results for the bound states of $T_{cc}$ and $T_{bb}$ with given $I(J^{P})$ and discuss the spin structures of 
the bound states obtained by our analyses. In Sec.~\ref{sec;summary}, we summarize our results and discussions.  

\section{Formalism}\label{sec;formalism}
\subsection{Lagrangian}\label{subsec;Lagrangian}
In the heavy quark limit (HQL), the heavy pseudoscalar and heavy vector mesons which include the heavy quarks are degenerate
due to HQS.
Therefore we define the heavy meson field $H_{a}$ 
written as the direct sum of 
the heavy pseudoscalar meson field $P_{a}$ and the heavy vector meson field $P^{*}_{a\mu}$ as follows~\cite{Neubert:1993mb,Grinstein:1995uv,Casalbuoni:1996pg}:
\begin{align}
  H_{a} &= \frac{1+\Slash{v}}{2}[P^{*}_{a\mu}\gamma^{\mu} - P_{a}\gamma_{5}]\label{eq;Ha},\\
  \bar{H}_{a} &= \gamma_{0}H_{a}^{\dagger}\gamma_{0} = [P^{*\dagger}_{a\mu}\gamma^{\mu} + P_{a}\gamma_{5}]\frac{1+\Slash{v}}{2}\label{eq;complex_Ha}.
\end{align}
$\bar{H}_{a}$ is the complex conjugate of $H_{a}$. Here, $v^{\mu}$ is the four velocity of the heavy quark which satisfies $v^2=1$, $v^{0}>0$ and $(1+\Slash{v})/2$ is the projective operator which projects out the positive-energy component in the heavy quark.
Also, the heavy meson filed $H_{a}$ is transformed as $H_{a}\to D(\Lambda)H_{a}D^{-1}(\Lambda)$ under the Lorentz transformation and $H_{a} \to S_{Q}H_{a}U^{\dagger}_{q}$ under 
the spin transformation for the heavy quark and the chiral transformation for a light quark.
Here $D(\Lambda)$ is the Lorentz transformation matrix, $S_{Q}$ is the spin transformation matrix and $U_{q}$ is the chiral transformation matrix.\\
\indent
We construct the interaction effective Lagrangian for a pseudoscalar meson~\cite{Casalbuoni:1996pg}:
\begin{equation}
  \L_{\pi} = ig\Tr[H_{b}\gamma_{\mu}\gamma_{5}\A_{ba}^{\mu}\bar{H}_{a}].
  \label{eq;L_ps}
\end{equation}
Here, $\A^{\mu}$ is the axial-vector current which is defined by   
\begin{equation}
  \A_{\mu} = \frac{1}{2} [\xi^{\dagger}(\partial_{\mu}\xi)-\xi(\partial_{\mu}\xi^{\dagger})],
\end{equation}
where $\xi$ is the nonlinear representation written as
\begin{equation}
  \xi = \exp\left(\frac{\vec{\tau}\cdot\vec{\pi}}{2f_{\pi}}\right)
\end{equation}
with the pion decay constant $f_{\pi} \simeq 93\,\mathrm{MeV}$. The pion field $\hat{\pi}$ is defined by 
\begin{equation}
  \hat{\pi} 
  = \frac{1}{\sqrt{2}}\begin{pmatrix}
    \pi^{0}&\sqrt{2}\pi^{+}\\ 
    \sqrt{2}\pi^{-}&-\pi^{0}
  \end{pmatrix}
  =\frac{1}{\sqrt{2}}
  \vec{\pi}\cdot\vec{\tau},\label{eq;pion_field}
\end{equation}
where $\vec{\tau}$ is the Pauli matrices. We obtain the interaction effective Lagrangians for the $\pi P^{(*)}P^{(*)}$ vertices by expanding Eq.~\eqref{eq;L_ps}~\cite{Ohkoda:2012hv}: 

\begin{align}
  \L_{\pi PP^{*}} &= -\frac{g}{f_{\pi}}(P^{*\dagger\mu}_{a}P_{b}+P^{\dagger}_{\mu}P^{*\mu}_{b})\partial_{\mu}\vec{\tau}\cdot\vec{\pi},\\
  \L_{\pi P^{*}P^{*}} &= i\frac{g}{f_{\pi}}\epsilon^{\mu\nu\alpha\beta}v_{\mu}P_{a\nu}^{*\dagger}P_{b\alpha}^{*}\partial_{\beta}\vec{\pi}\cdot\vec{\tau}.
\end{align}
However $\L_{\pi PP} = 0$ due to the parity conservation.

We can also obtain the interaction effective Lagrangians for vector mesons $v$ and a $\sigma$ meson~\cite{Casalbuoni:1996pg,Yamaguchi:2019vea}:
\begin{align}
   \L_{v} &= -i\beta\Tr[H_{b}v^{\mu}\rho_{\mu}\bar{H}_{a}] + i\lambda\Tr[H_{b}\sigma^{\mu\nu}{F_{\mu\nu}(\rho)}_{ba}\bar{H}_{a}],\\
  \L_{\sigma} &= g_{s}\Tr[H\sigma \bar{H}],
\end{align}
where  
\begin{align}
   F_{\mu\nu}(\rho) &= \partial_{\mu}\rho_{\nu} - \partial_{\nu}\rho_{\mu} + [\rho_{\mu},\rho_{\nu}],\\
  g_{V} &= \frac{m_{\rho}}{\sqrt{2}f_{\pi}},\ 
  \rho_{\mu} = \frac{ig_{V}}{\sqrt{2}}\hat{\rho}_{\mu}.
\end{align}
The vector meson fields $\hat{\rho}_{\mu}$ are defined by 
\begin{equation}
  \hat{\rho}_{\mu} = \frac{1}{\sqrt{2}}{
    \begin{pmatrix}
      \rho^{0}+\omega&\sqrt{2}\rho^{+}\\
      \sqrt{2}\rho^{-}&-\rho^{0}+\omega
    \end{pmatrix}
  }_{\mu} = \frac{1}{\sqrt{2}}(\vec{\tau}\cdot\vec{\rho}_{\mu} + \omega_{\mu}\boldsymbol{1}).
\end{equation}
\noindent
We calculate the Lagrangian for $v P^{(*)}P^{(*)}$ and $\sigma P^{(*)}P^{(*)}$ vertices~\cite{Ohkoda:2012hv}:
 \begin{align}
  \L_{v} &= \L_{vPP} + \L_{vPP^{*}} + \L_{vP^{*}P^{*}},\\
  \L_{\sigma} &= \L_{\sigma PP} + \L_{\sigma P^{*}P^{*}},
 \end{align}
where 
\begin{align}
  &\L_{vPP} = \sqrt{2}\beta g_{V}PP^{\dagger}v\cdot \hat{\rho},\\
  &\L_{vPP^{*}} = -2\sqrt{2}g_{V}\lambda\epsilon^{\mu\nu\alpha\beta}v_{\mu}(P^{*\dagger}_{\nu}P + P^{\dagger}P_{\nu})\partial_{\alpha}\hat{\rho}_{\beta},\\
  &\L_{vP^{*}P^{*}} = \sqrt{2}\beta g_{V}P\cdot P^{*\dagger}v\cdot\hat{\rho}
  + i2\sqrt{2}\lambda g_{V}P^{*\mu\dagger}P^{*\nu}(\partial_{\mu}\hat{\rho}_{\nu} - \partial_{\nu}\hat{\rho}),\\
  &\L_{\sigma PP} = -2g_{s}P^{\dagger}P\sigma,\\
  &\L_{\sigma P^{*}P^{*}} = 2g_{s}P^{*\dagger}_{\mu}P^{*\mu}\sigma.
\end{align}

\subsection{Hamiltonian}\label{subsec;Hamiltonian}
The OBEP is obtained by using these Lagrangians~\cite{Ohkoda:2012hv}:
\begin{itemize}
  \item $\pi$
    \begin{align}
      V^{\pi}_{PP^{*}-P^{*}P} &= \frac{1}{3}{\left(\frac{g}{2f_{\pi}}\right)}^2[
        -\vecsta{\varepsilon}\cdot\vec{\varepsilon}\, D(r;m_{\pi})
        +\vecsta{\varepsilon}\cdot\vec{\varepsilon}\, C(r;m_{\pi})
        +S_{\varepsilon^{*}\varepsilon}\, T(r;m_{\pi})
      ]
      \vec{\tau}_{1}\cdot\vec{\tau}_{2},\\
      V^{\pi}_{PP^{*}-P^{*}P^{*}} &= -\frac{1}{3}{\left(\frac{g}{2f_{\pi}}\right)}^2[
        -\vecsta{\varepsilon}\cdot\vec{T}\, D(r;m_{\pi})
        +\vecsta{\varepsilon}\cdot\vec{T}\, C(r;m_{\pi})
        +S_{\varepsilon^* T} \,T(r;m_{\pi})
      ]\vec{\tau}_{1}\cdot\vec{\tau}_{2},\\
      V^{\pi}_{PP-P^*P^*} &= \frac{1}{3}{\left(\frac{g}{2f_{\pi}}\right)}^2[
        -\vecsta{\varepsilon}\cdot\vecsta{\varepsilon}\, D(r;m_{\pi}) 
        + \vecsta{\varepsilon}\cdot\vecsta{\varepsilon}C(r;m_{\pi})
        + S_{\varepsilon^*\varepsilon^*}T(r;m_{\pi})
      ]\vec{\tau}_{1}\cdot\vec{\tau}_{2},\\
      V^{\pi}_{P^{*}P^{*}-P^{*}P^{*}} &= \frac{1}{3}{\left(\frac{g}{2f_{\pi}}\right)}^2[
        -\vec{T}\cdot\vec{T}\,D(r;m_{\pi}) 
        +\vec{T}\cdot\vec{T}\,C(r;m_{\pi})
        +S_{TT}\,T(r;m_{\pi})
      ]\vec{\tau}_{1}\cdot\vec{\tau}_{2},
    \end{align}
  \item vector mesons ($\rho,\ \omega$)
    \begin{align}
      V^{v}_{PP-PP} =& {\left(\frac{\beta g_{V}}{2m_{v}}\right)}^2C(r;m_{v})\vec{\tau}_{1}\cdot\vec{\tau}_{2},\\
      V^{v}_{PP^{*}-PP^{*}} =& {\left(\frac{\beta g_{V}}{2m_{v}}\right)}^2C(r;m_{v})\vec{\tau}_{1}\cdot\vec{\tau}_{2},\\
      V^{v}_{PP^{*}-P^{*}P} =& \frac{1}{3}{(\lambda g_{V})}^2[
        -2\vecsta{\varepsilon}\cdot\vec{\varepsilon}\,D(r;m_{v})
        +2\vecsta{\varepsilon}\cdot\vec{\varepsilon}\,C(r;m_{v})
        -S_{\varepsilon^{*}\varepsilon}\,T(r;m_{v})
      ]\vec{\tau}_{1}\cdot\vec{\tau}_{2},\\
      V^{v}_{PP^{*}-P^{*}P^{*}} =& -\frac{1}{3}{(\lambda g_{V})}^2[
        -2\vecsta{\varepsilon}\cdot\vec{T}\,D(r;m_{v})
        +2\vecsta{\varepsilon}\cdot\vec{T}\,C(r;m_{v})
        -S_{\varepsilon T}\,T(r;m_{v})
      ]\vec{\tau}_{1}\cdot\vec{\tau}_{2},\\
      V^{v}_{PP-P^*P^*} =& \frac{1}{3}{(\lambda g_{V})}^2[
        -2\vecsta{\varepsilon}\cdot\vecsta{\varepsilon}\,D(r;m_{v})
        +2\vecsta{\varepsilon}\cdot\vecsta{\varepsilon}\,C(r;m_v)
        -S_{\varepsilon^*\varepsilon^*}\,T(r;m_v)
      ]\vec{\tau}_1\cdot\vec{\tau}_2,\\
      V^{v}_{P^{*}P^{*}-P^{*}P^{*}} =& {\left(\frac{\beta g_{V}}{2m_{v}}\right)}^2C(r;m_{v})\vec{\tau}_{1}\cdot\vec{\tau}_{2}\notag\\
      &+\frac{1}{3}{(\lambda g_{V})}^2[
        -2\vec{T}\cdot\vec{T}\,D(r;m_{v})
        +2\vec{T}\cdot\vec{T}\, C(r;m_{v})
        -S_{TT}\,T(r;m_{v})
      ]\vec{\tau}_{1}\cdot\vec{\tau}_{2}
    \end{align}
  \item $\sigma$
    \begin{align}
      V^{\sigma}_{PP-PP} &= -{\left(\frac{g_{s}}{m_{\sigma}}\right)}^2C(r;m_{\sigma}),\\
      V^{\sigma}_{PP^{*}-PP^{*}} &= -{\left(\frac{g_{s}}{m_{\sigma}}\right)}^2C(r;m_{\sigma}),\\
      V^{\sigma}_{P^{*}P^{*}-P^{*}P^{*}} &= -{\left(\frac{g_{s}}{m_{\sigma}}\right)}^2C(r;m_{\sigma}),
    \end{align}
\end{itemize}
where we adopt a static approximation, which means that energy transfers are neglected. 
The potentials considering an energy transfer have been done in Ref.~\cite{Asanuma:2023atv}.
In the following, we ignore the $D(r)$ term, which implies the delta function, because we focus on the long-range and 
middle-range parts of the meson exchange forces. 
$\varepsilon^{\mu}$ is the polarization vector and $\vec{T}$ is the spin-one operator, which are defined by
\begin{align}
  \varepsilon^{(0)\mu} &= (0,0,0,1), \\ 
  \varepsilon^{(\pm)\mu} &= \frac{1}{\sqrt{2}}(0,1,\pm i,0), \\
  \vec{T} &= -i\vecsta{\varepsilon}\times\vec{\varepsilon},
\end{align}
respectively. 
$S_{12}(\hat{q}) = S_{\O_{1}\O_{2}}(\hat{q})$ is 
the tensor operator 
defined by
\begin{equation}
  S_{12}(\hat{q}) = S_{\O_{1}\O_{2}}(\hat{q})
  = 3(\vec{\O}_{1}\cdot\hat{q})(\vec{\O}_{2}\cdot\hat{q})
  -\vec{\O}_{1}\cdot\vec{\O}_{2},\label{eq;tensor_op}
\end{equation}
where $\hat{q}=\vec{q}/|\vec{q}|$. 
$C(r;m)$ and $T(r;m)$ are the central and the tensor potentials, which are defined by
\begin{align}
  C(r;m) &= \int\frac{d^3 q}{{(2\pi)}^3}\frac{m^2}{\vecsqu{q} + m^2}e^{i\vec{q}\cdot\vec{r}}{F(\vec{q};m)}^2,\label{eq;central_fun}\\
  S_{12}(r)T(r;m) &= \int\frac{d^3 q}{{(2\pi)}^3}S_{12}(\hat{q})\frac{-\vecsqu{q}}{\vecsqu{q} + m^2}e^{i\vec{q}\cdot\vec{r}}{F(\vec{q};m)}^2,\label{eq;tensor_fun}
\end{align}
respectively. Here we use 
\begin{equation}
  F(\vec{q};m) = \frac{\Lambda^2-m^2}{\Lambda^2 + \vecsqu{q}}\label{eq;form_factor}
\end{equation}
as the form factor $F(\vec{q};m)$ in order to consider a hadron size, where $\Lambda$ is a cutoff parameter.
Inserting this form factor into Eqs.~\eqref{eq;central_fun} and \eqref{eq;tensor_fun},
we obtain the central and the tensor functions:
\begin{align}
  C(r;m) = & \frac{m^2}{4\pi}\left[
    \frac{e^{-mr}}{r} - \frac{e^{-\Lambda r}}{r} - \frac{\Lambda^2-m^2}{2\Lambda}e^{-\Lambda r}
  \right]\label{eq;central_fun_explicit},\\
  T(r;m) = &
  \frac{1}{4\pi}(3 + 3mr + m^2r^2
 )\frac{e^{-mr}}{r^3}
  -\frac{1}{4\pi}(3 + 3\Lambda r + \Lambda^2r^2)\frac{e^{-\Lambda r}}{r^3}\notag\\
  &+\frac{1}{4\pi}\frac{m^2 - \Lambda^2}{2}(1+\Lambda r)\frac{e^{-\Lambda r}}{r}.
\end{align}

We solve the Schr\"{o}dinger equation in order to obtain the binding energy, wave functions and mixing ratios. Then, let us show the Hamiltonian below:
\begin{equation}
  H_{I(J^{P})} = K_{I(J^{P})} + \sum_{\mathrm{boson}=\pi,\rho,\omega,\sigma}V^{\HM}_{\mathrm{boson},I(J^{P})},\label{eq;Hamiltonian}
\end{equation} 
where $K_{I(J^{P})}$ is the kinetic energy and $V^{\HM}_{\mathrm{boson},I(J^P)}$ is the OBEP in HMB.

\section{Numerical results}\label{sec;numerical_results}
In this section, we show the numerical results. First, we discuss the doubly charmed tetraquark $T_{cc}$.
The cutoff parameter $\Lambda$ is determined to reproduce the empirical binding energy of $T_{cc}$ with $0(1^+)$.  
We study the properties of the $0(1^+)$ state and also the possible existence of the bound states of $T_{cc}$ with other quantum numbers. 
Second, we discuss the doubly bottom tetraquark $T_{bb}$ with given $I(J^P)$. Finally, we consider $T_{QQ}$ in 
the HQL
in order to see the HQS multiplets of the doubly heavy tetraquark. 
In Table~\ref{table;param}, we show the masses of mesons and the parameters used in this research. 
Here, the value of the coupling constant of a $\sigma$ meson $g_s$ is uncertainly. In this work, we use $g_{s}=3.4$, which is determined to be one-third of the coupling constant
of a nucleon and a 
$\sigma$ meson~\cite{CLEO:2001foe}. 
We also vary the value of $g_{s}$ 
and discuss the $g_s$ dependence of results.

\renewcommand{\arraystretch}{1.25}
\begin{table}[htb]
  \centering\caption{The masses of mesons and the parameters~\cite{Workman:2022ynf,CLEO:2001foe,PhysRevD.99.094018,Isola:2003fh}.}\begin{tabular}{cc|cc}
    \toprule[0.3mm]
    masses &&parameters&\\\hline
    $m_{\pi}$&$138\,\mathrm{MeV}$\ \ \ &\ \ $g$&$0.59$\\
    $m_{\rho}$&$770\,\mathrm{MeV}$\ \ \ &\ \ $g_{V}$&$\frac{m_{\rho}}{\sqrt{2}f_{\pi}}$\\
    $m_{\omega}$&$782\,\mathrm{MeV}$\ \ \ &\ \ $\beta$&$0.9$\\
    $m_{\sigma}$&$500\,\mathrm{MeV}$\ \ \ &\ \ $\lambda$&$0.56\,\mathrm{GeV}^{-1}$\\
    $m_{D}$&$1868\,\mathrm{MeV}$\ \ \ &\ \ $g_{s}$&$3.4$\\
    $m_{D^{*}}$&$2009\,\mathrm{MeV}$\ \ \ &\ \ &\\
    $m_{B}$&$5279\,\mathrm{MeV}$\ \ \ &\ \ &\\
    $m_{B^{*}}$&$5325\,\mathrm{MeV}$\ \ \ &\ \ &\\\bottomrule[0.3mm]
  \end{tabular}
  \label{table;param}
\end{table}
\renewcommand{\arraystretch}{1}

\subsection{Doubly charmed tetraquark $T_{cc}$}\label{subsec;Tcc}
\subsubsection{$I(J^{P})=0(1^{+})$}\label{sebsec;Tcc_01+}
\begin{figure}[tb]
  \centering
  \includegraphics[scale=0.5]{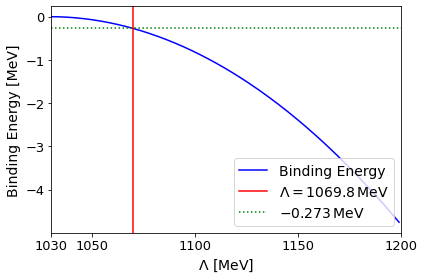}
  \caption{The binding energy of $T_{cc}$ with $0(1^{+})$ for $\Lambda=1030-1200\,\mathrm{MeV}$. 
 The vertical 
 solid line indicates
 $\Lambda=1069.8\,\mathrm{MeV}$ and 
  the horizontal 
 dotted line 
 does
 $-0.273\,\mathrm{MeV}$, which is the experimental value of the $T_{cc}$ binding energy~\cite{LHCb:2021vvq}.}\label{fig;Tcc_BE_Lambda}
\end{figure}

\begin{figure}[tb]
  \includegraphics[scale=0.5]{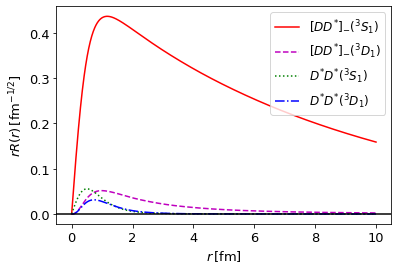}
  \caption{The wave function of each channel for $T_{cc}$ with $0(1^{+})$.
 The solid, dashed, dotted and dashed-dotted lines show the wave functions of ${[DD^{*}]}_{-}({}^3S_{1})$, ${[DD^{*}]}_{-}({}^3D_{1})$, $D^{*}D^{*}({}^3S_{1})$ and $D^{*}D^{*}({}^3D_{1})$ channels, respectively.}
  \label{fig;Tcc_wavefunction}
\end{figure}

\begin{figure*}[t]
  \centering
  \begin{tabular}{cc}
    (a) $\pi$ & (b) $\rho$\\
    \includegraphics[scale=0.5]{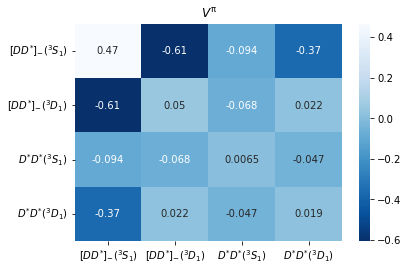}&
    \includegraphics[scale=0.5]{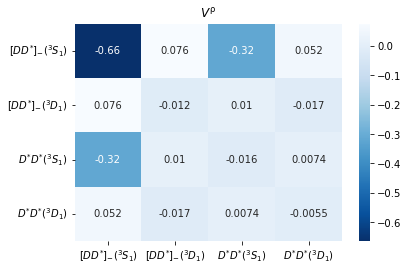}\\
    (c) $\omega$ & (d) $\sigma$\\
    \includegraphics[scale=0.5]{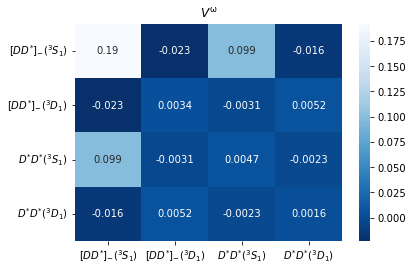}&
    \includegraphics[scale=0.5]{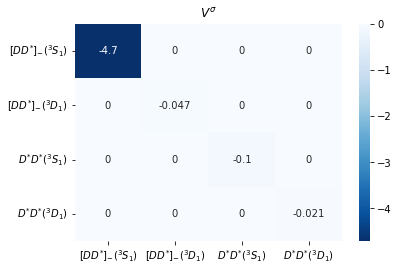}
  \end{tabular}
  \caption{
    The expectation values of the OBEP for $T_{cc}$ with $0(1^{+})$. The value is given in units of MeV. 
  }\label{fig;T_cc_Potential}
\end{figure*}

As mentioned above, the pseudoscalar and vector mesons are degenerate in the HQL. 
Here, we consider the heavy pseudoscalar meson $D$ and the heavy vector meson $D^{*}$, both of which include a charm quark. The mass difference between 
$D$ and $D^{*}$ is approximately $140\,\mathrm{MeV}$. Then, we consider $D$ and $D^{*}$ as the HQS doublet.  
Thus 
the threshold energies of $DD,\ DD^{*}$ and $D^{*}D^{*}$ 
channels are approximately degenerate. 
We consider these channels coupled by the one boson exchange interactions. 
Let us show the possible channels 
for the $I(J^P)=0(1^+)$ state 
$\psi^{\mathrm{HM}}_{0(1^{+})}$:
\begin{equation}
  \psi^{\mathrm{HM}}_{0(1^{+})} = \begin{pmatrix}
    \ket{{[PP^{*}]}_{-}({}^3S_{1})}\\\ket{{[PP^{*}]}_{-}(^3D_{1})}\\\ket{P^{*}P^{*}({}^3S_{1})}\\\ket{P^{*}P^{*}({}^3D_{1})}
  \end{pmatrix},
\end{equation}
where we use the notations ${[PP^{*}]}_{\pm}=\frac{1}{\sqrt{2}}(PP^{*}\pm P^{*}P)$.\\
Also, we calculate the kinetic energy $K_{0(1^{+})}$ and potential energies $V^{\mathrm{HM}}_{\text{boson},0(1^{+})}$: 
\renewcommand{\arraystretch}{1.25}
\begin{align}
  K_{0(1^{+})} &= \diag\left(
    -\frac{1}{2\mu_{PP^{*}}}\triangle_{0},
    -\frac{1}{2\mu_{PP^{*}}}\triangle_{2},
    -\frac{1}{2\mu_{P^{*}P^{*}}}\triangle_{0} + \Delta m_{PP^{*}},
    -\frac{1}{2\mu_{P^{*}P^{*}}}\triangle_{2} + \Delta m_{PP^{*}}
    \right),\\
  V^{\mathrm{HM}}_{\pi,0(1^{+})} 
    &=\begin{pmatrix}
      -C_{\pi}&\!\!\sqrt{2}T_{\pi}&\!\!2C_{\pi}&\!\!\sqrt{2}T_{\pi}\\\vspace{1mm}
      \sqrt{2}T_{\pi}&\!\!-C_{\pi}-T_{\pi}&\!\!\sqrt{2}T_{\pi}&\!\!2C_{\pi}-T_{\pi}\\\vspace{1mm}
      2C_{\pi}&\!\!\sqrt{2}T_{\pi}&\!\!-C_{\pi}&\!\!\sqrt{2}T_{\pi}\\\vspace{1mm}
      \sqrt{2}T_{\pi}&\!\!2C_{\pi}-T_{\pi}&\!\!\sqrt{2}T_{\pi}&\!\!-C_{\pi}-T_{\pi}
    \end{pmatrix}\label{eq;Vpi_01+},\\
    V^{\mathrm{HM}}_{v,0(1^{+})}
    &=\begin{pmatrix}
      C^{\prime}_{v}-2C_{v}&\!\!-\sqrt{2}T_{v}&\!\!4C_{v}&\!\!-\sqrt{2}T_{v}\\\vspace{1mm}
      -\sqrt{2}T_{v}&\!\!C^{\prime}_{v}-2C_{v}+T_{v}&\!\!-\sqrt{2}T_{v}&\!\!4C_{v}+T_{v}\\\vspace{1mm}
      4C_{v}&\!\!-\sqrt{2}T_{v}&\!\!C^{\prime}_{v}-2C_{v}&\!\!-\sqrt{2}T_{v}\\\vspace{1mm}
      -\sqrt{2}T_{v}&\!\!4C_{v}+T_{v}&\!\!-\sqrt{2}T_{v}&\!\!C^{\prime}_{v}-2C_{v}+T_{v}
    \end{pmatrix}\label{eq;Vv_01+},\\
  V^{\mathrm{HM}}_{\sigma,0(1^{+})} &= \begin{pmatrix}
    C_{\sigma}&0&0&0\\\vspace{1mm}
    0&C_{\sigma}&0&0\\\vspace{1mm}
    0&0&C_{\sigma}&0\\\vspace{1mm}
    0&0&0&C_{\sigma}
  \end{pmatrix}\label{eq;Vsigma_01+},
\end{align}
where 
\begin{align}
  \mu_{P^{(*)}P^{(*)}} &= \frac{m_{P^{(*)}}m_{P^{(*)}}}{m_{P^{(*)}} + m_{P^{(*)}}},\\
  \triangle_{l} &= \frac{d^2}{dr^2} - \frac{l(l+1)}{r^2},\\
  \Delta m_{PP^{*}} &= m_{P^{*}} - m_{P},\\
  C_{\pi} &= \frac{1}{3}{\left(\frac{g}{2f_{\pi}}\right)}^2 C(r;m_{\pi})\vec{\tau}_{1}\cdot\vec{\tau}_{2},\\
  T_{\pi} &= \frac{1}{3}{\left(\frac{g}{2f_{\pi}}\right)}^2 T(r;m_{\pi})\vec{\tau}_{1}\cdot\vec{\tau}_{2},\\
  C^{\prime}_{v} &= {\left(\frac{\beta g_{V}}{2m_{v}}\right)}^2 C(r;m_{v})\vec{\tau}_{1}\cdot\vec{\tau}_{2},\\
  C_{v} &= \frac{1}{3}{(\lambda g_{V})}^2 C(r;m_{v})\vec{\tau}_{1}\cdot\vec{\tau}_{2},\\
  T_{v} &= \frac{1}{3}{(\lambda g_{V})}^2 T(r;m_{v})\vec{\tau}_{1}\cdot\vec{\tau}_{2},\\
  C_{\sigma} &= -{\left(\frac{g_s}{m_{\sigma}}\right)}^2 C(r;m_{\sigma}).
\end{align}
\renewcommand{\arraystretch}{1}
For a omega meson, we remove $\vec{\tau}_{1}\cdot \vec{\tau}_{2}$ because the isospin of a omega meson is $0$. \\
\indent
First, we consider only the one $\pi$ exchange force in the $D^{(\ast)}D^{(\ast)}$ molecule.
In this case, no bound state of $T_{cc}$ with $0(1^{+})$ exists for reasonable $\Lambda$. 
However considering the OBEP as the interaction of a hadronic molecule, we obtain 
the bound state of 
$T_{cc}$ with $0(1^{+})$ as shown Fig.~\ref{fig;Tcc_BE_Lambda}. 
The binding energy of $T_{cc}$ reported by LHCb,
$0.273\,\mathrm{MeV}$~\cite{LHCb:2021vvq}, is obtained for $\Lambda = 1069.8\,\mathrm{MeV}$. 
The bound state properties with $\Lambda = 1069.8\,\mathrm{MeV}$, the wave functions, the mixing ratios and the root-mean square distance (RMS), 
are shown in Fig.~\ref{fig;Tcc_wavefunction} and Table~\ref{table;Tcc_mixing}. 
Here, the mixing ratios $f$ and the RMS are defined by 
\begin{align*}
  f(\text{channel}) &= \braket{\chi_{\mathrm{channel}}|\chi_{\mathrm{channel}}},\\
  \sqrt{\braket{r^2}} &= \sqrt{\braket{\chi|r^2|\chi}},
\end{align*}
where $\chi = rR(r)$ and $\chi_{\mathrm{channel}}$ is the wave function of a channel. 
Also, the potential expectation values of the OBEP are shown in Fig.~\ref{fig;T_cc_Potential} as like matrices. 
Each component in Fig.~\ref{fig;T_cc_Potential} shows 
the expectation values of corresponding potential component in Eqs.~\eqref{eq;Vpi_01+}-\eqref{eq;Vsigma_01+}. 
For example, $0.47$ in Fig.~\ref{fig;T_cc_Potential}~(a) is the expectation value of the 
$(1,1)$
component of $V^\pi$.
Our analyses show that 
considering only 
the OPEP cannot earn the attractive force necessary to bind $T_{cc}$. However, we get the bound state of 
$T_{cc}$ in the case of OBEP. In fact, Fig.~\ref{fig;T_cc_Potential}~(d) shows that the $(1,1)$ component of $V^\sigma$
generates 
the strongest attraction
among all potential expectation values, indicating that the exchange of a $\sigma$ meson is the most significant.
Also, the $(1,2)$, $(2,1)$, $(1,4)$ and $(4,1)$ components of the OPEP 
tensor term 
in Fig.~\ref{fig;T_cc_Potential}~(a) and the $(1,1)$, $(1,3)$ and $(3,1)$ components of the $\rho$ exchange potential in Fig.~\ref{fig;T_cc_Potential}~(b) are important 
to produce an attraction. 
Thus, 
the bound state of $T_{cc}$ with $0(1^{+})$ 
are obtained 
in the case of the OBEP. 

\renewcommand{\arraystretch}{1.25}
\begin{table}[tbp]
  \caption{The mixing ratios of each channel and the RMS for $T_{cc}$ with $0(1^{+})$ where $\Lambda=1069.8\,\mathrm{MeV}$.}
  \centering\begin{tabular}{cc}
    \toprule[0.3mm]
    mixing ratio and RMS &\\\hline
    ${[DD^{*}]}_{-}({}^3S_{1})$ & $99.2\,\%$\\
    ${[DD^{*}]}_{-}({}^3D_{1})$ & $0.467\,\%$\\
    $D^{*}D^{*}({}^3S_{1})$ & $0.229\,\%$\\
    $D^{*}D^{*}({}^3D_{1})$ & $0.0854\,\%$\\
    RMS&$6.43\,\mathrm{fm}$\\\bottomrule[0.3mm]
  \end{tabular}\label{table;Tcc_mixing}
\end{table}
\renewcommand{\arraystretch}{1}

\begin{figure}
  \includegraphics[scale=0.5]{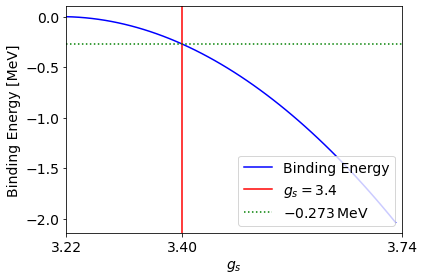}
  \caption{The relation of the binding energy of $T_{cc}$ and $g_{s}$. 
 The vertical 
 solid
 line 
 indicates
 $g_{s}=3.4$ and the horizontal 
 dotted
 line 
 does
 $-0.273\,\mathrm{MeV}$ 
  which is the experimental value.}\label{fig;Tcc_gs_BE}
\end{figure}

\begin{figure}
  \includegraphics[scale=0.5]{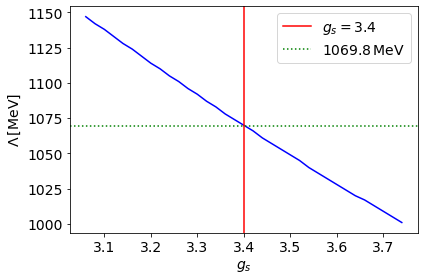}
  \caption{The relation 
 between $g_s$ and
 $\Lambda$ which 
 reproduces
 the experimental value 
 of the $T_{cc}$ binding energy.  The vertical solid and horizontal dotted lines indicate $g_{s}=3.4$ and $\Lambda=1069.8$ MeV, respectively. }
 \label{fig;Tcc_gs_Lambda}
\end{figure}

\indent
Next, we 
vary the value of 
the sigma coupling $g_{s}$ by $\pm 10\,\%$ 
because of the uncertainty of $g_s$, and study 
$g_{s}$ dependence of the binding energy 
for $\Lambda=1069.8\,\mathrm{MeV}$, as shown Fig.~\ref{fig;Tcc_gs_BE}.  
This analysis shows that we can obtain the bound state of $T_{cc}$ for $g_{s} \geq 3.22$ and the binding energy of $T_{cc}$ for $\Lambda=1069.8\,\mathrm{MeV}$ varies greatly 
when we 
vary $g_s$.
This analysis implies the binding energy highly depends on $g_{s}$.  
Futhermore, we also
investigate 
$g_{s}$ dependence of the cutoff parameter $\Lambda$ 
to reproduce the experimental binding energy of $T_{cc}$ as shown in Fig.~\ref{fig;Tcc_gs_Lambda}. 
This result shows that even if $g_s$ is varied as $3.06\leq g_s\leq 3.74$, we obtain the bound state with the empirical binding energy having the reasonable cutoff as $1001.3\text{\, MeV}\leq\Lambda\leq 1147.1$\text{\, MeV}.

\subsubsection{Other quantum numbers}\label{subsec;Tcc_other}
In this section, we 
study
$T_{cc}$ with other quantum numbers,
$0(0^{-})$, $0(1^{-})$, $1(0^{+})$, $1(0^{-})$, $1(1^{+})$, and $1(1^{-})$. 
The possible channels of these quantum numbers 
are shown in Table~\ref{table;other_channel} and the potential matrices are shown in appendix~\ref{appendix;H}. 
Here, we consider the case where $g_{s} = 3.06, 3.4, 3.74$. 
We 
solve
the Schr\"{o}dinger equations for these quantum numbers, but 
we 
obtain
no bound states of $T_{cc}$ with other quantum numbers, as shown in Table~\ref{table;T_cc_other_gs}.  
\renewcommand{\arraystretch}{1.25}
\begin{table*}[tb]
  \caption{The possible channels for each quantum number. We use the notation ${[PP^*]}_\pm = \frac{1}{\sqrt{2}}(PP^*\pm P^*P)$ and $^{2S+1}L_{J}$ where $S$ is the total spin, $L$ is the orbital angular momentum and $J$ is the total angular momentum ~\cite{Ohkoda:2012hv}.}\label{table:possible_channel}
  \centering
  \large
   \begin{tabular}{c|c}
    \toprule[0.3mm]
    $I(J^{P})$ & channels \\\hline
    $0(0^{-})$ & ${[PP^{*}]}_{+}({}^3P_{0})$\\[5pt]
    $0(1^{+})$ & ${[PP^{*}]}_{-}({}^3S_{1}),{[PP^{*}]}_{-}({}^3D_{1}),P^{*}P^{*}({}^3S_{1}),P^{*}P^{*}({}^3D_{1})$\\[5pt]
    $0(1^{-})$ & $PP({}^1P_{1}),{[PP^{*}]}_{+}({}^3P_{1}),P^{*}P^{*}({}^1P_{1}),P^{*}P^{*}({}^5P_{1}),P^{*}P^{*}({}^5F_{1})$\\[5pt]
    $1(0^{+})$ & $PP({}^1S_{0}),P^{*}P^{*}({}^1S_{0}),P^{*}P^{*}({}^5D_{0})$\\[5pt]
    $1(0^{-})$ & ${[PP^{*}]}_{-}({}^3P_{0}),P^{*}P^{*}({}^3P_{0})$\\[5pt]
    $1(1^{+})$ & ${[PP^{*}]}_{+}({}^3S_{1}),{[PP^{*}]}_{+} ({}^3D_{1}),P^{*}P^{*}({}^5D_{1})$\\[5pt]
    $1(1^{-})$ & ${[PP^{*}]}_{-}({}^3P_{1}),P^{*}P^{*}({}^3P_{1})$\\\bottomrule[0.3mm]
   \end{tabular}\label{table;other_channel}
\end{table*}
\begin{table}[hbt]
  \centering
  \caption{The binding energies  ($B$)
 of $T_{cc}$ with given $I(J^{P})$.
 In the table, $-B$ is displayed in units of MeV.
The dependence of the binding energy on $\Lambda$ and $g_{s}$ is shown.  
The set of $(g_s, \Lambda)$
 is determined to reproduce 
the experimental value of $T_{cc}$ 
 as shown in Fig.~\ref{fig;Tcc_gs_Lambda}.
 }\label{table;T_cc_other_gs}
  \begin{tabular}{cccc}
    \toprule[0.3mm]
    $g_{s}$\, &\, $3.06$\, &\, $3.4$\, &\, $3.74$\, \\
    $\Lambda[\mathrm{MeV}]$\, &\, $1147.1$\, &\, $1069.8$\, &\, $1001.3$\\\hline
    $0(0^{-})$\, &\, -\, &\, -\, &\, -\\
    $0(1^{+})\, $&\, $-0.273$\, &\, $-0.273$\, &\, $-0.273$\, \\
    $0(1^{-})\, $&\, -\, &\, -\, &\, -\\
    $1(0^{+})\, $&\, -\, &\, -\, &\, -\\
    $1(0^{-})\, $&\, -\, &\, -\, &\, -\\
    $1(1^{+})\, $&\, -\, &\, -\, &\, -\\
    $1(1^{-})\, $&\, -\, &\, -\, &\, -\\\bottomrule[0.3mm]
  \end{tabular}
\end{table}
\renewcommand{\arraystretch}{1}

\subsection{Doubly bottom tetraquark $T_{bb}$}\label{subsec;Tbb}
In this section, we 
consider $T_{bb}$ 
as a hadronic molecule of $B$ and $B^\ast$, including two bottom quarks. 
In Sec.~\ref{subsec;Tcc}, we considered $D$ and $D^{*}$ as a HQS doublet because the mass difference of $D$ and $D^{*}$ 
is smaller than those in the light quark sectors. 
Similarly, we think of $B$ and $B^{*}$ as a HQS doublet since the mass difference of $B$ and $B^{*}$ is approximately $45\,\mathrm{MeV}$, and the $BB^\ast-B^\ast B^\ast$ channel-coupling effect is expected to be enhanced. 

\subsubsection{$I(J^{P}) = 0(1^{+})$}\label{subsubsec;Tbb_01+}
\begin{figure}
  \includegraphics[scale=0.5]{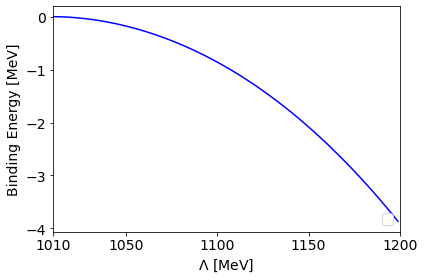}
  \caption{The relation between the binding energy of $T_{bb}$ with $0(1^{+})$ and $\Lambda$ for OPEP.}\label{fig;Tbb_pi_BE}
\end{figure}

\begin{figure}[tb]
  \includegraphics[scale=0.5]{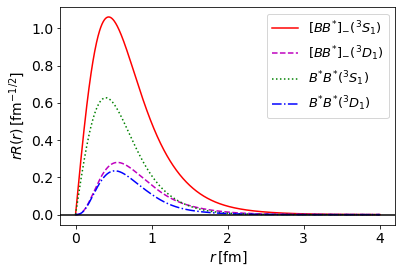}
  \caption{The wave function of each channel for $T_{bb}$ with $0(1^{+})$.
 The solid, dashed, dotted and dashed-dotted lines show the wave functions of ${[BB^{*}]}_{-}({}^3S_{1})$, ${[BB^{*}]}_{-}({}^3D_{1})$, $B^{*}B^{*}({}^3S_{1})$ and $B^{*}B^{*}({}^3D_{1})$ channels, respectively.
 }\label{fig;Tbb_wavefucntion}
\end{figure}

\begin{figure*}[t]
  \centering
  \begin{tabular}{cc}
    (a) $\pi$ & (b) $\rho$\\
    \includegraphics[scale=0.5]{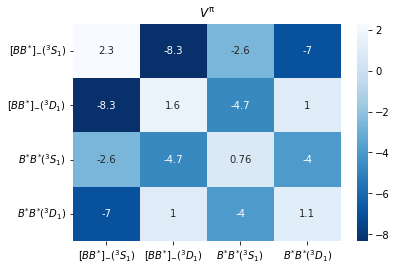}\label{fig;T_bb_component_pi_01plus} &
    \includegraphics[scale=0.5]{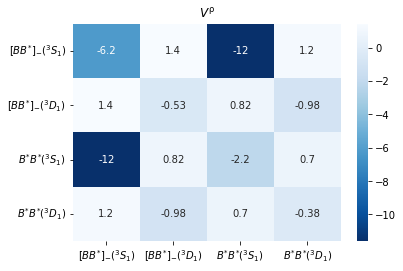}\label{fig;T_bb_component_rho_01plus}\\
    (c) $\omega$ & (d) $\sigma$\\
    \includegraphics[scale=0.5]{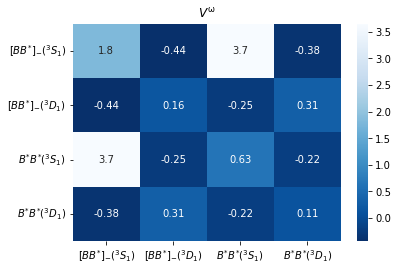} \label{fig;T_bb_component_omega_01plus}&
    \includegraphics[scale=0.5]{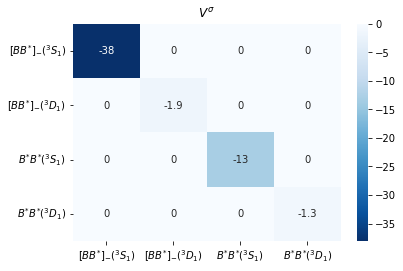}\label{fig;T_bb_component_sigma_01plus}
  \end{tabular}
  \caption{
    The expectation values of the OBEP for $T_{bb}$ with $0(1^{+})$. The value is given in units of MeV. 
  }\label{fig;T_bb_Potential}
\end{figure*}

First, we discuss the $T_{bb}$ with $0(1^{+})$ as a hadronic molecule whose interaction is only $\pi$ exchange force. 
By solving the Schr\"odinger equation of the $B^{(\ast)}B^{(\ast)}$ two-body system, 
we obtain
the bound state for $T_{bb}$ with $0(1^{+})$,
while $T_{cc}$ does not bind only by the OPEP with the reasonable cutoff as discussed in the previous section. 
The binding energies with various cutoff $\Lambda$ are shown in Fig.~\ref{fig;Tbb_pi_BE}, where the binding energy of $T_{bb}$ increases as $\Lambda$ increases. 
The analysis only with the OPEP indicates that it is highly likely that the bound state of $T_{bb}$ exists because additional attractions from the short-range forces are also expected as discussed below.  
\\
\indent
We also consider the case where the interaction of a hadronic molecule is the OBEP. 
We use the same cutoff of $T_{cc}$, $\Lambda=1069.8\,\mathrm{MeV}$,
determined in the previous section and 
calculate the binding energy of $T_{bb}$ with $0(1^{+})$. 
As a result, we 
find that the binding energy 
is $46.0\,\mathrm{MeV}$ and also 
get wave functions, mixing ratios and the RMS as shown in Fig.~\ref{fig;Tbb_wavefucntion} and Table~\ref{table;Tbb_mixing}. 
These results show that the channel of $B^{*}B^{*}({}^3S_{1})$ is important in addition to the one of ${[BB^{*}]}_{-}(^3S_1)$
unlike $T_{cc}$. 
The importance of the $B^{*}B^{*}({}^3S_{1})$ channel is also seen in the potential expectation values in Fig.~\ref{fig;T_bb_Potential}.
The expectation values of the one $\sigma$ exchange potential in Fig.~\ref{fig;T_bb_Potential}~(d) shows that the dominant component is given by the $(1,1)$ component, which is the same in the case of $T_{cc}$ as shown in Fig.~\ref{fig;T_cc_Potential}~(d).
In addition, the $(3,3)$ component in Fig.~\ref{fig;T_bb_Potential}~(d) also produces the strong attraction, whereas the corresponding component of $T_{cc}$ is not important. 
The other meson exchanges also have an important role to generate the attraction.
The (1,2), (2,1), (1,4) and (4,1) components in Fig.~\ref{fig;T_bb_Potential}~(a) are important, which implies the tensor force of the one pion exchange contributes to bind $T_{bb}$ just like the case of $T_{cc}$. 
The (1,3) and (3,1) components in Fig.~\ref{fig;T_bb_Potential}~(b) primarily contributed by the $\rho$ meson exchange are important to generate the attraction. 
These are because the mass difference of $BB^{*}$ and $B^{*}B^{*}$ is 
smaller than that of $DD^{*}$ and $D^{*}D^{*}$. Therefore the channels ${[BB^{*}]}_{-}$ and $B^{*}B^{*}$ are more coupled with each other. 
However, we emphasize that
similar to $T_{cc}$, the $\sigma$ meson exchange potential
has the dominant contribution to earn
the attractive force binding the $T_{bb}$. 

\renewcommand{\arraystretch}{1.25}
\begin{table}[tbp]
  \caption{The mixing ratios of each channel and RMS for $T_{bb}$ with $0(1^{+})$ where $\Lambda=
 1069.8
 \,\mathrm{MeV}$.}
  \centering\begin{tabular}{cc}
    \toprule[0.3mm]
    mixing ratio and RMS &\\\hline
    ${[BB^{*}]}_{-}({}^3S_{1})$ & $70.7\,\%$\\
    ${[BB^{*}]}_{-}({}^3D_{1})$ & $4.71\,\%$\\
    $B^{*}B^{*}({}^3S_{1})$ & $21.6\,\%$\\
    $B^{*}B^{*}({}^3D_{1})$ & $3.00\,\%$\\
    RMS&$0.620\,\mathrm{fm}$\\\bottomrule[0.3mm]
  \end{tabular}\label{table;Tbb_mixing}
\end{table}
\renewcommand{\arraystretch}{1}

\subsubsection{Other quantum numbers}
In this section, we 
discuss $T_{bb}$ with given $I(J^{P})$.  
The obtained binding energies are summarized in Table~\ref{table;T_bb_other_gs}.
As a result, 
many bound states of $T_{bb}$ appear. 
Table~\ref{table;T_bb_other_gs} shows that the $(g_s,\Lambda)$ dependence of the $T_{bb}$ binding energies, where the parameter set $(g_s,\Lambda)$ is determined to reproduce the empirical binding energy of $T_{cc}$ as shown in Table.~\ref{table;T_cc_other_gs} and Fig.~\ref{fig;Tcc_gs_Lambda}. 
Examining the $(g_s,\Lambda)$ dependence is useful to understand natures of the different isospin states.
Table~\ref{table;T_bb_other_gs} shows that as $g_{s}$ increases 
and simultaneously $\Lambda$ decreases 
(the $\sigma$ exchange is enhanced while the other meson exchanges are suppressed), 
the bound states  
of $T_{bb}$ with the isospin $I=0$ 
become shallower, while 
the ones with the isospin $I=1$ becomes deeper.  
The difference between the isospon states is caused by the isospin factor $\vec{\tau}_1\cdot \vec{\tau}_2$ of the isovector $\pi,\rho$ exchange potentials: 
\begin{equation*}
  \vec{\tau}_{1}\cdot\vec{\tau}_{2}=\left\{
    \begin{array}{l}
      -3\  \text{for}\ I=0\\
      1\ \  \ \,\text{for}\ I=1
    \end{array}
  \right. \, .
\end{equation*}
Since the strength of the $\vec{\tau}_1\cdot \vec{\tau}_2$ factor of the $I=0$ channel is three times larger than the one of the $I=1$ channel, the $\pi,\rho$ exchange potentials have non-negligible role in the $I=0$ bound states, while it is suppressed in the case of $I=1$. 
It can also be seen in the $(g_s,\Lambda)$ dependence of the energy expectation values in Table~\ref{table;T_bb_com_other}.
For $0(1^+)$, the off-diagonal components are drastically reduced as $\Lambda$ decreases, which are mainly contributed by the isovector $\pi,\rho$ exchange potentials.
The reduction of the expectation value is also seen for the $0(0^-)$ state.
On the other hand, the expectation values of the $I=1$ bound states are more sensitive to the change of $g_s$, being the coupling constant of the $\sigma$ exchange in the diagonal components.

\renewcommand{\arraystretch}{1.25}
\begin{table}[tb]
  \centering
  \caption{The binding energies 
 of $T_{bb}$ with given $I(J^{P})$.
 The same convention as Table~\ref{table;T_cc_other_gs} is used.  }\label{table;T_bb_other_gs}
  \begin{tabular}{cccc}
    \toprule[0.3mm]
    $g_{s}$\, &\, $3.06$\, &\, $3.4$\, &\, $3.74$\, \\
    $\Lambda[\mathrm{MeV}]$\, &\, $1147.1$\, &\, $1069.8$\, &\, $1001.3$\\\hline
    $0(0^{-})$\, &\, $-30.7$\, &\, $-24.4$\, &\, $-19.2$\\
    $0(1^{+})\, $&\, $-56.2$\, &\, $-46.0$\, &\, $-37.9$\, \\
    $0(1^{-})\, $&\, -\, &\, -\, &\, -\\
    $1(0^{+})\, $&\, $-3.70$\, &\, $-7.23$\, &\, $-10.8$\\
    $1(0^{-})\, $&\, -\, &\, -\, &\, -\\
    $1(1^{+})\, $&\, $-0.0254$\, &\, $-2.46$\, &\, $-6.98$\\
    $1(1^{-})\, $&\, -\, &\, -\, &\, -\\\bottomrule[0.3mm]
  \end{tabular}
\end{table}
\renewcommand{\arraystretch}{1}

\renewcommand{\arraystretch}{1.25}
\begin{table}[bt]  
  \centering\caption{ 
 Dependence of 
 the energy expectation value of 
  $T_{bb}$ for given $I(J^{P})$ on the parameters $(g_{s},\Lambda)$. 
 Only the important components are shown in the table.
  The expectation value is given in units of MeV. 
 }\begin{tabular}{ccccc}  
    \toprule[0.3mm] 
    &$g_{s}$\, &\, $3.06$\, &\, $3.40$\, &\, $3.74$\, \\
    &$\Lambda[\mathrm{MeV}]$\, &\, $1147.1$\, &\, $1069.8$\, &\, $1001.3$\\ 
    \midrule[0.3mm]
    $0(1^{+})$&
    $(1,1)$\, &\, $-39$\, &\, $-40$\, &\, $-41$\\ \cline{2-5}
    & $(1,2)$\, &\, \multirow{2}{*}{$-8.6$}\, &\, \multirow{2}{*}{$-7.3$}\, &\, \multirow{2}{*}{$-6.2$}\\
    & $(2,1)$\, &\\ \cline{2-5}
    & $(1,3)$\, &\, \multirow{2}{*}{$-16$}\, &\, \multirow{2}{*}{$-11$}\, &\, \multirow{2}{*}{$-6.7$}\\
    & $(3,1)$\, &\\ \cline{2-5}
    & $(1,4)$\, &\, \multirow{2}{*}{$-7.4$}\, &\, \multirow{2}{*}{$-6.2$}\, &\, \multirow{2}{*}{$-5.1$}\\
    & $(4,1)$\, &\\ \cline{2-5}
    & $(2,3)$\, &\, \multirow{2}{*}{$-5.5$}\, &\, \multirow{2}{*}{$-4.1$}\, &\, \multirow{2}{*}{$-2.9$}\\
    & $(3,2)$\, &\\ \cline{2-5}
    & $(3,4)$\, &\, \multirow{2}{*}{$-4.8$}\, &\, \multirow{2}{*}{$-3.5$}\, &\, \multirow{2}{*}{$-2.4$}\\
    & $(4,3)$\, &\\ \cline{2-5}
    & $(3,3)$\, &\, $-17$\, &\, $-14$\, &\, $-10$\\
    \midrule[0.3mm]
    $0(0^{-})$& 
    $V_{\mathrm{total}}\, $&\, $-108$\, &\, $-90.6$\, &\, $-75.7$\\
    \midrule[0.3mm]
    $1(0^{+})$& 
    $(1,1)$\, &\, $-13$\, &\, $-22$\, &\, $-30$\\ \cline{2-5}
    & $(1,2)$\, &\, \multirow{2}{*}{$-2.6$}\, &\, \multirow{2}{*}{$-1.8$}\, &\, \multirow{2}{*}{$-0.99$}\\
    & $(2,1)$\, &\\ \cline{2-5}
    & $(1,3)$\, &\, \multirow{2}{*}{$-0.55$}\, &\, \multirow{2}{*}{$-0.74$}\, &\, \multirow{2}{*}{$-0.85$}\\
    & $(3,1)$\, &\\ \cline{2-5}
    & $(2,2)$\, &\, $-1.9$\, &\, $-1.3$\, &\, $-0.72$\\
    \midrule[0.3mm]
    $1(1^{+})$& 
    $(1,1)$\, &\, $-0.72$\, &\, $-10$\, &\, $-21$\\ \cline{2-5}
    & $(1,2)$\, &\, \multirow{2}{*}{$-0.057$}\, &\, \multirow{2}{*}{$-0.4$}\, &\, \multirow{2}{*}{$-0.51$}\\
    & $(2,1)$\, &\\ \cline{2-5}
    & $(1,3)$\, &\, \multirow{2}{*}{$-0.095$}\, &\, \multirow{2}{*}{$-0.76$}\, &\, \multirow{2}{*}{$-1.1$}\\
    & $(3,1)$\, &\\
    \bottomrule[0.3mm]
  \end{tabular}
  \label{table;T_bb_com_other}
\end{table}
\renewcommand{\arraystretch}{1}
\subsection{Doubly heavy tetraquarks $T_{QQ}$ in the heavy quark limit}\label{subsec;TQQ}
\subsubsection{Light-Cloud Basis}
Until now, we have discussed  
the hadronic molecules of $P$ and $P^{*}$ as $T_{cc}$ and $T_{bb}$.
In this section, by introducing the light-cloud basis (LCB), we consider the HQS multiplet structure of the molecules in 
the HQL.
As discussed in Refs.~\cite{Yasui:2013vca,Yamaguchi:2014era,Shimizu:2019ptd}, HQS and LCB are useful to classify bound states by the heavy quark spin 
and total angular momentum of the light cloud.

We can obtain the LCB by implementing the unitary transformation to the hadronic-molecule basis (HMB)~\cite{Yasui:2013vca,Yamaguchi:2014era,Shimizu:2019ptd}:
\begin{align}
  {\bigg[L\ {\Big[{\big[S_{Q_{1}}S_{q_{1}}\big]}_{S_{1}}{\big[S_{Q_{2}}S_{q_{2}}\big]}_{S_{2}}\Big]}_{S}\bigg]}_{J}\ \to\ 
  {\bigg[ {\Big[ S_{Q_{1}}S_{Q_{2}}\Big]}_{S_{Q}}\ {\Big[L\ {\big[S_{q_{1}}S_{q_{2}}\big]}_{S_{q}}\Big]}_{J_{l}} \bigg]}_{J},\label{eq;LCB}
\end{align}
where 
$L$ is the orbital momentum, 
$S_{Q_i}$ and $S_{q_i}$ ($i=1,2$) are the spins of the heavy quark $Q_i$ and light antiquark $q_i$ of the heavy meson $P_i^{(\ast)}=Q_i\bar{q}_i$ with the spin $S_i$, respectively. 
$S$ and $J$ are the total spin and angular momentum of two heavy mesons, 
$S_Q$ and $S_q$ are the spins of the heavy diquark and the light anti-diquark, and $J_{l}$ is the spin of 
the light cloud. 
This transformation leads to find the spin structures including the diquark spins inside the hadronic molecule,  
which we cannot see in the HMB.

Here, we implement this unitary transformation from the HMB to the LCB 
for $0(1^{+})$ as an example. 
First, 
the transformation from the wavefunction in the HMB, $\psi^{\HM}_{0(1^{+})}$, to that in the LCB, $\psi^{\LC}_{0(1^{+})}$, is given by
\begin{align}
  \psi^{\mathrm{LC}}_{0(1^{+})} 
  &= U^{-1}_{0(1^{+})} \psi^{\mathrm{HM}}_{0(1^{+})}\notag \\
  &= 
  \begin{pmatrix}
    {\Ket{{\Big[{\big[QQ\big]}_{1}\ {\big[S\ {[\bar{q}\bar{q}]}_{0}\big]}_{0}\Big]}_{1}}}\vspace{1mm}\\
    {\Ket{{\Big[{\big[QQ\big]}_{0}\ {\big[S\ {[\bar{q}\bar{q}]}_{1}\big]}_{1}\Big]}_{1}}}\vspace{1mm}\\
    {\Ket{{\Big[{\big[QQ\big]}_{0}\ {\big[D\ {[\bar{q}\bar{q}]}_{1}\big]}_{1}\Big]}_{1}}}\vspace{1mm}\\
    {\Ket{{\Big[{\big[QQ\big]}_{1}\ {\big[D\ {[\bar{q}\bar{q}]}_{0}\big]}_{2}\Big]}_{1}}}\vspace{1mm}
  \end{pmatrix}, 
 \label{eq:LCB-0(1+)}
 \\
  U_{0(1^{+})} &= \begin{pmatrix}
    -\frac{1}{\sqrt{2}}&\frac{1}{\sqrt{2}}&0&0\vspace{1mm}\\
    0&0&\frac{1}{\sqrt{2}}&-\frac{1}{\sqrt{2}}\vspace{1mm}\\
    \frac{1}{\sqrt{2}}&\frac{1}{\sqrt{2}}&0&0\vspace{1mm}\\
    0&0&\frac{1}{\sqrt{2}}&\frac{1}{\sqrt{2}}
  \end{pmatrix},
\end{align}
where $U_{0(1^{+})}$ is the unitary matrix determined by the Clebsh-Gordan coefficient. 
Second, 
using $U_{0(1^{+})}$,
we transform the potential matrices $V^{\HM}_{\text{boson},0(1^{+})}$ in the HMB to $V^{\LC}_{\text{boson},0(1^{+})}$ in the LCB.
 Then we obtain
the block-diagonal potential matrices:
\begin{align} 
  V^{\mathrm{LC}}_{\pi,0(1^{+})} &= U^{-1}_{0(1^{+})}V^{\mathrm{HM}}_{\pi,0(1^{+})}U_{0(1^{+})}\notag\\
  &=\left(\begin{array}{c|cc|c}
    -3C_{\pi}&0&0&0\\
    \hline
    0&C_{\pi}&2\sqrt{2}T_{\pi}&0\\
    0&2\sqrt{2}T_{\pi}&C_{\pi}-2T_{\pi}&0\\
    \hline
    0&0&0&-3C_{\pi}
  \end{array}\right),\\
  V^{\mathrm{LC}}_{v,0(1^{+})} &= U^{-1}_{0(1^{+})}V^{\mathrm{HM}}_{v,0(1^{+})}U_{0(1^{+})}\notag\\
  &=\left(
    \begin{array}{c|cc|c}
      C^{\prime}_{v}-6C_{v}&0&0&0\\\hline
      0&C^{\prime}_{v}+2C_{v}&-2\sqrt{2}T_{v}&0\\
      0&-2\sqrt{2}T_{v}&C^{\prime}_{v}+2C_{v}+2T_{v}&0\\\hline
      0&0&0&C^{\prime}_{v}-6C_{v}
    \end{array}
  \right),\\
  V^{\mathrm{LC}}_{\sigma,0(1^{+})} &= U^{-1}_{0(1^{+})}V^{\mathrm{HM}}_{\sigma,0(1^{+})}U_{0(1^{+})}\notag\\
  &=\left(
    \begin{array}{c|cc|c}
      C_{\sigma}&0&0&0\\\hline
      0&C_{\sigma}&0&0\\
      0&0&C_{\sigma}&0\\\hline
      0&0&0&C_{\sigma}
    \end{array}
  \right).
\end{align}
Since states with different $S_Q$ or $J_l$ are decoupled in the HQL, the off-diagonal components mixing these states vanish in the LCB. 
Hence if a bound state is obtained, it is an eigenstate corresponding to one of the components which is characterized by $S_Q$ and $J_l$. Thus 
these $\psi^{\LC}_{I(J^{P})}$ and matrices $V^{\LC}_{\text{boson},0(1^{+})}$ enable us to find the spin structures of the diquark.

We note that the matrix elements of the block-diagonal potential coincide with those of the meson exchange potential between corresponding light quarks. 
In this study, we employ only the light meson exchange interactions which work between light quarks and the rest heavy quark is a spectator. Thus using the transformation 
from the HMB to LCB, the potentials between heavy mesons are rewritten as those between the light quarks inside the heavy mesons.

The diquark spins $S_Q$ and $J_l$ also indicate the HQS multiplet structure of bound states. For instance, the first component of Eq.~\eqref{eq:LCB-0(1+)} has $S_Q=1$ and $J_l=0$. 
Since only $J=1$ can be generated from $S_Q=1$ and $J_l=0$, the bound state for this component should belong to the HQS singlet and hence it has no HQS partner. 
Similarly the coupled channel system of the second and third components 
has $S_Q=0$ and $J_l=1$. Thus the bound state is in the HQS singlet. 
Therefore, from the LCB of the $0(1^+)$ state, we find that if there is a $S$-wave bound state, it belongs to the HQS singlet. 
On the other hand, the fourth component having $S_Q=1$ and $J_l=2$ with $L=2$ may have the HQS partners with high $J$ because both $S_Q$ and $J_l$ are nonzero. 

We also show an example of 
the unitary transformation
for $0(0^-)$ and $0(1^-)$, 
where we can see the HQS multiplet structure of them. As seen in the case of $0(1^+)$ states, the light-cloud basis, $\psi^\LC_{0(0^-)}$ and $\psi^\LC_{0(1^-)}$, and 
the one boson exchange potential matrices, $V^\LC_{\mathrm{boson},0(0^-)}$ and $V^\LC_{\mathrm{boson},0(1^-)}$ can be obtained under the unitary transformations as
\begin{align}
  \psi^{\mathrm{LC}}_{0(0^{-})} &= \begin{pmatrix}
    -\Ket{{\Big[{\big[QQ\big]}_{1}\ {\big[P\ {[\bar{q}\bar{q}]}_{1}\big]}_{1}\Big]}_{0}}
  \end{pmatrix},\\
  \psi^{\mathrm{LC}}_{0(1^{-})} &= U^{-1}_{0(1^{-})}\psi^{\mathrm{HM}}_{0(1^{-})}\notag\\
    &= \begin{pmatrix}
      \Ket{{\Big[{\big[QQ\big]}_{0}\ {\big[P\ {[\bar{q}\bar{q}]}_{0}\big]}_{1}\Big]}_{1}}\vspace{1mm}\\
      \Ket{{\Big[{\big[QQ\big]}_{1}\ {\big[P\ {[\bar{q}\bar{q}]}_{1}\big]}_{0}\Big]}_{1}}\vspace{1mm}\\
      \Ket{{\Big[{\big[QQ\big]}_{1}\ {\big[P\ {[\bar{q}\bar{q}]}_{1}\big]}_{1}\Big]}_{1}}\vspace{1mm}\\
      \Ket{{\Big[{\big[QQ\big]}_{1}\ {\big[P\ {[\bar{q}\bar{q}]}_{1}\big]}_{2}\Big]}_{1}}\vspace{1mm}\\
      \Ket{{\Big[{\big[QQ\big]}_{1}\ {\big[F\ {[\bar{q}\bar{q}]}_{1}\big]}_{2}\Big]}_{1}}
    \end{pmatrix},\\
  V^\LC_{\pi,0(0^-)} &= (C_{\pi}+2T_{\pi}),\label{eq:VpiLCB00-}\\
  V^{\mathrm{LC}}_{\pi,0(1^{-})} &= U^{-1}_{0(1^-)}V^\HM_{\pi,0(1^-)}U_{0(1^-)}\notag\\
  &= \left(\begin{array}{c|c|c|cc}
    -3C_{\pi}&0&0&0&0\\
    \hline
    0&C_{\pi}-4T_{\pi}&0&0&0\\
    \hline
    0&0&C_{\pi}+2T_{\pi}&0&0\\
    \hline
    0&0&0&C_{\pi}-\frac{2}{5}T_{\pi}&\frac{6\sqrt{6}}{5}T_{\pi}\\
    0&0&0&\frac{6\sqrt{6}}{5}T_{\pi}&C_{\pi}-\frac{8}{5}T_{\pi}
  \end{array}\right) ,\label{eq:VpiLCB01-}
\end{align}
where only the one pion exchange potentials are shown as an example. Comparing the potential matrices in Eqs.~\eqref{eq:VpiLCB00-} and \eqref{eq:VpiLCB01-} shows
the same components, $C_\pi+2T_\pi$, where their spin structures are also the same, $S_Q=1$ and $J_l=1$.
It is also found for the other one boson exchange matrices. 
Thus the eigenstates of the corresponding Hamiltonian component in the $0(0^-)$ and $0(1^-)$ states are degenerate in the HQL,
and belong to the HQS multiplet. In addition,
since a combination of $S_Q=1$ and $J_l=1$ generates
$J=0,1,2$, 
i.e. $S_Q\otimes J_l=1\otimes 1=0\oplus 1\oplus 2$,
we also expect that the $0(2^-)$ state has the same component, and the corresponding components of $0(0^-)$, $0(1^-)$ and $0(2^-)$ belong to the same HQS triplet. \\

We also obtain the other $\psi^{\LC}_{I(J^{P})}$ and $V^{\LC}_{\text{boson},I(J^{P})}$ summarized in appendix~\ref{appendix;LCB}.

\subsubsection{$T_{QQ}$ in HQL and the spin structure}

We calculate the binding energy of $T_{QQ}$ with given $I(J^{P})$ in the HQL. 
In fact, the Schr\"odinger equations for $T_{QQ}$ cannot be solved numerically, because the reduced mass of the two mesons diverge. To demonstrate a computation in the HQL, we take 
$m_{P}=m_{P^{*}}=5m_{B^{*}}$ 
which implies that the masses of the pseudoscalar meson and the vector meson are degenerate. Using these masses, 
we discuss the HQS multiplet structure for $T_{cc}$.
These results are shown in Table~\ref{table;HL_01plus}~-~\ref{table;HL_11minus}.

For the $0(1^{+})$ state in the HQL, 
the result summarized in Table~\ref{table;HL_01plus} shows that there are five bound states. 
However
the origin of these states is different, which is indicated by 
their mixing ratios.  
The mixing ratios $f$ of the ground state, and 
first, third and fourth excited states in the HMB are 
\begin{align*}
  f({[PP^{*}]}_{-}({}^3S_{1})):f(P^{*}P^{*}({}^3S_{1})) &= 1:1,\\
  f({[PP^{*}]}_{-}({}^3D_{1})):f(P^{*}P^{*}({}^3D_{1})) &= 1:1,
\end{align*}
obtained as the $S$-$D$ mixing states. These relations show that these bound states are composed of 
\begin{align}
 \begin{pmatrix}
    {\Ket{{\Big[{\big[QQ\big]}_{0}\ {\big[S\ {[\bar{q}\bar{q}]}_{1}\big]}_{1}\Big]}_{1}}}\vspace{1mm}\\
    {\Ket{{\Big[{\big[QQ\big]}_{0}\ {\big[D\ {[\bar{q}\bar{q}]}_{1}\big]}_{1}\Big]}_{1}}}\vspace{1mm}\\
 \end{pmatrix}
\end{align}
components in Eq.~\eqref{eq:LCB-0(1+)} in the LCB. Thus, we also find that the diquark spins of these bound states are $S_Q=0$ and $S_q=1$.
On the other hand, 
the mixing ratio of the remaining state, the second excited state, is 
\begin{align*}
  f({[PP^{*}]}_{-}({}^3S_{1}))=f(P^{*}P^{*}({}^3S_{1})) &= 50\,\%,\\
  f({[PP^{*}]}_{-}({}^3D_{1}))=f(P^{*}P^{*}({}^3D_{1})) &= 0\,\%,
\end{align*}
having no $D$-wave state. Therefore this bound state is built by the ${\Ket{{\Big[{\big[QQ\big]}_{1}\ {\big[S\ {[\bar{q}\bar{q}]}_{0}\big]}_{0}\Big]}_{1}}}$ component in Eq.~\eqref{eq:LCB-0(1+)} 
in the LCB, where the diquark spins are obtained by $S_{Q}=1$ and $S_{q}=0$.

\begin{figure}[htb]
  \centering
  \includegraphics[scale=0.5]{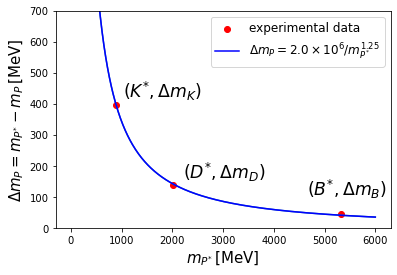}
  \caption{
 Heavy vector meson mass dependence of $\Delta m_{P} = m_{P^{*}}-m_{P}$. 
 The dots 
 are the experimental data of $K^{*}$, $D^{*}$, and $B^{*}$ from left to right. 
  The solid line 
 is a re-fitting result referring to Ref.~\cite{Yamaguchi:2011xb}, and this result is $\Delta m_P = 2.00\times 10^{\,6}/{m_{P^{*}}}^{1.25}$. 
 }
\label{fig;mass_delta}
\end{figure}

\begin{figure*}
  \centering
  \includegraphics[scale=0.5]{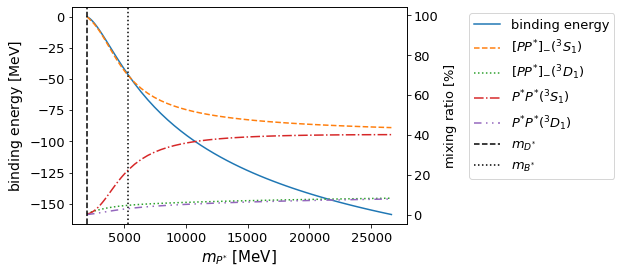}
  \caption{The mass dependence of the binding energy 
 and mixing ratios
 of $T_{QQ}$ with $0(1^+)$ for the grand state. 
  The horizontal axis shows the mass of $P^\ast$. 
 The vertical dashed and dotted lines indicate $m_{P^\ast}=m_{D^\ast}$ and $m_{P^\ast}=m_{B^\ast}$, respectively.
  }\label{fig;mass_E_P_01plus}
\end{figure*}

\begin{figure*}[t]
  \centering
 \includegraphics[scale=0.5]{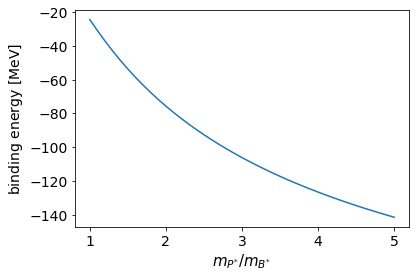}
  \caption{
    The mass dependence of the binding energy of $T_{QQ}$ with $0(0^{-})$ for the ground state.
 The horizontal axis shows the ratio between the mass of the heavy vector meson $P^*$ and that of $B^*$. 
  }\label{fig;mass_E_P_00minus}
\end{figure*}

\begin{figure*}[t]
  \centering
 \includegraphics[scale=0.5]{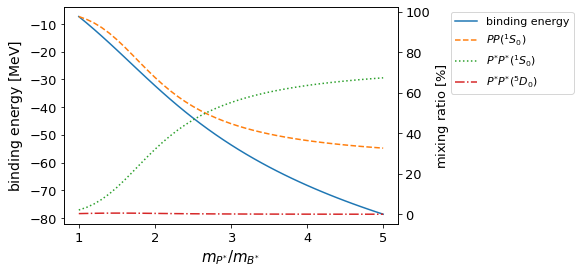}
  \caption{
    The mass dependence of the binding energy of $T_{QQ}$ with $1(0^{+})$ for the ground state.
 The same convention as Fig.~\ref{fig;mass_E_P_00minus} is used.
  }\label{fig;mass_E_P_10plus}
\end{figure*}

\begin{figure*}[t]
  \centering
 \includegraphics[scale=0.5]{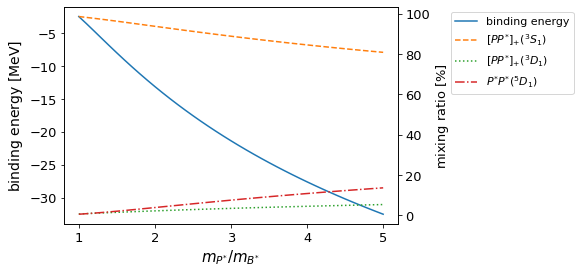}
  \caption{
    The mass dependence of the binding energy and mixing rations of $T_{QQ}$ with $1(1^{+})$ for the ground state.
 The same convention as Fig.~\ref{fig;mass_E_P_00minus} is used.
  }\label{fig;mass_E_P_11plus}
\end{figure*}

These results are obtained in the HQL, while in experiments, it is possible to observe doubly heavy tetraquarks with finite quark mass. Next, by reducing the heavy meson masses toward to the bottom and charmed meson regions, 
we connect the results of $T_{QQ}$ in the HQL, $T_{bb}$ and $T_{cc}$ where the HQS is not held exactly in the finite quark mass region.
Fig.~\ref{fig;mass_delta} shows the heavy vector meson mass dependence of $\Delta m_P = m_{P^{*}} - m_{P}$. 
By fitting the experimental data of the meson masses, we obtain $\Delta m_P=2.00\times 10^6/m_{P^*}^{1.25}$ as a function of $m_{P^\ast}$~\cite{Yamaguchi:2011xb}.
By using this 
function, 
we 
obtain 
the mass dependence of the energy eigenvalue and mixing ratios of $T_{QQ}$ with each quantum number, 
as shown in 
Figs.~\ref{fig;mass_E_P_01plus}-\ref{fig;mass_E_P_11plus}. 
The curves in these figures are continuous, thus we can see the origin of $T_{cc}$ and $T_{bb}$ in the HQL. 
As for $0(1^{+})$, Fig.~\ref{fig;mass_E_P_01plus} shows that the origin of $T_{cc}$ and $T_{bb}$ bound states obtained in this paper 
is the ground state of $T_{QQ}$
in the HQL with the spin structures $S_{Q}=0,S_{q}=1$,
indicating that this $T_{QQ}$ state belongs to the HQS singlet.   
Thus the $T_{cc}$ state reported by LHCb is originated from the HQS singlet state in the HQL, and the HQS partner is not present.

In Ref.~\cite{Meng:2020knc}, the author analyzed the doubly heavy tetraquark by using the quark model and found two bound states of $T_{bb}$ with $0(1^+)$. One is the deeply bound state and the other is the shallow one. 
The author considered the difference in the two internal structures: 
The deeply bound state 
has a very compact structure, while the shallow one is a molecular state. When the author changed the bottom quark to charm or strange quark 
for the deeply bound state, this binding energy decreased in order of the reduced masses of the diquarks. 
This behavior was explained 
by the color electric force which provides attraction for the color $\bar{\mathbf{3}}$ $QQ$ diquark. 
This color structure indicates that the deeply bound state in Ref.~\cite{Meng:2020knc} has the spin structure $S_Q=1, S_q=0$ because of the Fermi-Dirac statistics.
However, in our analysis
$T_{cc}$ and $T_{bb}$ with $0(1^+)$ having the spin structures $S_{Q}=0$, $S_{q}=1$ are obtained as the grand state, which means these states contain the color $\mathbf{6}$ $QQ$ diquark. 
The color electric force does not provide an attraction in the color $\mathbf{6}$ $QQ$ diquark,
while 
the tensor force of the meson exchange does.
As mentioned in Sec.~\ref{sebsec;Tcc_01+} and \ref{subsubsec;Tbb_01+}, 
the tensor force of the OPEP is important. 
The strong tensor force prefers the spin structures $S_{Q}=0$, $S_q=1$ as the origins of $T_{cc}$ and $T_{bb}$ with $0(1^+)$.

For $T_{QQ}$ with $0(0^{-})$ and $0(1^{-})$ in the HQL,
Tables~\ref{table;HL_00minus}~and~\ref{table;HL_01minus} show that every bound state of $T_{QQ}$ with $0(0^{-})$ is degenerate with a certain bound state of $T_{QQ}$ with $0(1^{-})$ because of the HQS. 
As an example, the ground state of $T_{QQ}$ with $0(0^{-})$ is degenerate with that of $0(1^{-})$. 
In fact, $V^{\LC}_{\text{boson},0(0^{-})}$ and $V^{\LC}_{\text{boson},0(1^{-})}$ have the same component 
, and hence these bound states belong to the same HQS multiplet.
Fig.~\ref{fig;mass_E_P_00minus}  
shows that the $T_{bb}$ bound state for $0(0^-)$ continuously connects to the ground state of $T_{QQ}$ in the HQL, having the spin structure $(S_{Q},S_{q},J_l)=(1,1,1)$. Thus the origin of the $T_{bb}(0(0^-))$ bound state should belong to the HQS triplet, 
where $0(1^-)$ and $0(2^-)$ states are present to be the HQS partners. However, in our analysis no bound state for these quantum numbers is found even for the bottom quark mass region. We expect that these states are found as a resonance above the thresholds.
The resonances with $0(1^-)$ and $0(2^-)$ have been discussed in literature~\cite{Ohkoda:2012hv,Meng:2021yjr}.

We also study the $T_{QQ}$ states for the isotriplet channel. For $1(0^{+})$ and $1(1^{+})$ states,
Tables~\ref{table;HL_10plus} and~\ref{table;HL_11plus} show that every bound state of $T_{QQ}$ 
with $1(1^{+})$ is degenerate with a certain bound state of $T_{QQ}$ with $1(0^{+})$. 
For the ground states in the HQL, their spin structures are $(S_{Q},S_{q},J_l)=(0,0,0)$ for $1(0^{+})$ and $(S_Q,S_q,J_l)=(1,1,1)$ for $1(1^{+})$. 
These spin structures indicate that the $1(0^{+})$ 
and $1(1^{+})$ ground states belong to the HQS singlet and triplet, respectively.
However, the $T_{QQ}(1(1^{+}))$ 
ground state is degenerate with the first excited state of $T_{QQ}(1(0^{+}))$ as shown in Tables~\ref{table;HL_10plus} and~\ref{table;HL_11plus}.
Figs.~\ref{fig;mass_E_P_10plus} and \ref{fig;mass_E_P_11plus} 
show that the $T_{QQ}$ ground states continuously connects to the $T_{bb}$ bound states. In Table~\ref{table;T_bb_other_gs}, the binding energies of $T_{bb}(1(0^+))$ and $T_{bb}(1(1^+))$ are similar. However, as found in the $T_{QQ}$, 
the origins of these $T_{bb}$ bound states in the HQL are different.

Finally we study the $1(0^{-})$ and $1(1^{-})$ states. We could not find a bound state for $T_{cc}$ and $T_{bb}$, while some bound states are obtained in the HQL. Tables~\ref{table;HL_10minus} and~\ref{table;HL_11minus} show that every bound state of $T_{QQ}$ with $1(1^{-})$ is degenerate 
with a certain bound state of $T_{QQ}$ with $1(0^{-})$. The ground states have the same spin structure 
$(S_{Q},S_{q},J_l)=(1,0,1)$,  
indicating that $T_{QQ}(1(0^{-}))$ and $T_{QQ}(1(1^{-}))$ are in the same HQS triplet. The remaining state should exist in $1(2^{-})$.

\renewcommand{\arraystretch}{1.25}
\begin{table*}[htb]
  \centering\caption{Energy eigenvalues $E$ ($=-B$ with binding energies $B$) 
 and mixing ratios of each channel for $0(1^{+})$ in 
 the HQL
 ($m_{P}=m_{P^{*}}=5m_{B^{*}}$). 
  The energy is given in units of MeV.}
 \begin{tabular}{cccccc}
    \toprule[0.3mm]
    &$E$ [MeV]&${[PP^{*}]}_{-}({}^3S_{1})$&${[PP^{*}]}_{-}({}^3D_{1})$&$P^{*}P^{*}({}^3S_{1})$&$P^{*}P^{*}({}^3D_{1})$\\
    \hline
    ground&$-162$&$41.9\,\%$&$8.07\,\%$&$41.9\,\%$&$8.07\,\%$\\
    1st&$-77.4$&$38.9\,\%$&$11.1\,\%$&$38.9\,\%$&$11.1\,\%$\\
    2nd&$-25.9$&$50.0\,\%$&$0\,\%$&$50.0\,\%$&$0\,\%$\\
    3rd&$-25.4$&$37.2\,\%$&$12.8\,\%$&$37.2\,\%$&$12.8\,\%$\\
    4th&$-3.07$&$37.3\,\%$&$12.7\,\%$&$37.3\,\%$&$12.7\,\%$\\\bottomrule[0.3mm]
 \end{tabular}\label{table;HL_01plus}
\end{table*}

\begin{table*}[htb]
  \centering\caption{Energy eigenvalues $E$ ($=-B$ with binding energies $B$) for $0(0^{-})$ in the HQL 
 ($m_{P}=m_{P^{*}}=5m_{B^{*}}$).
  The energy is given in units of MeV.}\begin{tabular}{cc}
    \toprule[0.3mm]
    &$E$ [MeV]\\
    \hline
    ground&$-141$\\
    1st&$-60.1$\\
    2nd&$-15.6$\\
    3rd&$-0.796$\\\bottomrule[0.3mm]
  \end{tabular}\label{table;HL_00minus}    
  \centering\caption{Energy eigenvalues $E$ ($=-B$ with binding energies $B$) and mixing ratios of each channel for $0(1^{-})$ in the 
 HQL 
 ($m_{P}=m_{P^{*}}=5m_{B^{*}}$).
  The energy is given in units of MeV.}
  \begin{tabular}{ccccccc}
    \toprule[0.3mm]
    &$E$ [MeV]&$PP({}^1P_{1})$&${[PP^{*}]}_{+}({}^3P_{1})$&$P^{*}P^{*}({}^1P_{1})$&$P^{*}P^{*}({}^5P_{1})$&$P^{*}P^{*}({}^5F_{1})$\\
    \hline
    ground&$-141$&$25.0\,\%$&$25.0\,\%$&$8.33\,\%$&$41.7\,\%$&$0\,\%$\\
    1st&$-104$&$32.4\,\%$&$32.4\,\%$&$10.8\,\%$&$2.16\,\%$&$22.3\,\%$\\
    2nd&$-60.1$&$25.0\,\%$&$25.0\,\%$&$8.33\,\%$&$41.7\,\%$&$0\,\%$\\
    3rd&$-38.7$&$30.4\,\%$&$30.4\,\%$&$10.1\,\%$&$2.03\,\%$&$27.1\,\%$\\
    4th&$-15.6$&$25.0\,\%$&$25.0\,\%$&$8.33\,\%$&$41.7\,\%$&$0\,\%$\\
    5th&$-6.40$&$29.8\,\%$&$29.8\,\%$&$9.93\,\%$&$1.99\,\%$&$28.5\,\%$\\
    6th&$-4.37$&$25.0\,\%$&$0\,\%$&$75.0\,\%$&$0\,\%$&$0\,\%$\\
    7th&$-0.796$&$25.0\,\%$&$25.0\,\%$&$8.33\,\%$&$41.7\,\%$&$0\,\%$\\\bottomrule[0.3mm]
  \end{tabular}\label{table;HL_01minus}
\end{table*}

\begin{table*}[htb]
  \centering\caption{Energy eigenvalues $E$ ($=-B$ with binding energies $B$) and mixing ratios of each channel for $1(0^{+})$ in the 
 HQL
 ($m_{P}=m_{P^{*}}=5m_{B^{*}}$).
  The energy is given in units of MeV.}
  \begin{tabular}{ccccc}
    \toprule[0.3mm]
    &$E$ [MeV]&$PP({}^1S_{0})$&$P^{*}P^{*}({}^1S_{0})$&$P^{*}P^{*}({}^5D_{0})$\\
    \hline
    ground&$-87.0$&$25.0\,\%$&$75.0\,\%$&$0\,\%$\\
    1st&$-33.4$&$58.5\,\%$&$19.5\,\%$&$22.1\,\%$\\
    2nd&$-21.8$&$25.0\,\%$&$75.0\,\%$&$0\,\%$\\
    3rd&$-7.61$&$43.0\,\%$&$14.3\,\%$&$42.6\,\%$\\
    4th&$-0.561$&$25.0\,\%$&$75.0\,\%$&$0\,\%$\\\bottomrule[0.3mm]
  \end{tabular}
  \label{table;HL_10plus}
  \centering\caption{Energy eigenvalues $E$ ($=-B$ with binding energies $B$) and mixing ratios of each channel for $1(1^{+})$ in the 
 HQL
 ($m_{P}=m_{P^{*}}=5m_{B^{*}}$).
  The binding energy is given in units of MeV.}
  \begin{tabular}{ccccc}
    \toprule[0.3mm]
    &$E$ [MeV]&${[PP^{*}]}_{+}({}^3S_{1})$&${[PP^{*}]}_{+}({}^3D_{1})$&$P^{*}P^{*}({}^5D_{1})$\\
    \hline
    ground&$-33.4$&$77.9\,\%$&$5.52\,\%$&$16.5\,\%$\\
    1st&$-7.61$&$57.4\,\%$&$10.7\,\%$&$32.0\,\%$\\\bottomrule[0.3mm]
  \end{tabular}
  \label{table;HL_11plus}
\end{table*}

\begin{table*}[htb]
  \centering\caption{Energy eigenvalues $E$ ($=-B$ with binding energies $B$) and mixing ratios of each channel for $1(0^{-})$ in the 
 HQL
 ($m_{P}=m_{P^{*}}=5m_{B^{*}}$).
  The binding energy is given in units of MeV.}
  \begin{tabular}{cccc}
    \toprule[0.3mm]
    &$E$ [MeV]&${[PP^{*}]}_{-}({}^3P_{0})$&$P^{*}P^{*}({}^3P_{0})$\\
    \hline
    ground&$-42.5$&$50.0\,\%$&$50.0\,\%$\\
    1st&$-33.9$&$50.0\,\%$&$50.0\,\%$\\
    2nd&$-6.09$&$50.0\,\%$&$50.0\,\%$\\
    3rd&$-4.27$&$50.0\,\%$&$50.0\,\%$\\\bottomrule[0.3mm]
  \end{tabular}
  \label{table;HL_10minus}
  \centering\caption{Energy eigenvalues $E$ ($=-B$ with binding energies $B$) and mixing ratios of each channel for $1(1^{-})$ in the 
 HQL ($m_{P}=m_{P^{*}}=5m_{B^{*}}$).
  The binding energy is given in units of MeV.}
  \begin{tabular}{cccc}
    \toprule[0.3mm]
    &$E$ [MeV]&${[PP^{*}]}_{-}({}^3P_{1})$&$P^{*}P^{*}({}^3P_{1})$\\
    \hline
    ground&$-42.5$&$50.0\,\%$&$50.0\,\%$\\
    1st&$-4.27$&$50.0\,\%$&$50.0\,\%$\\\bottomrule[0.3mm]
  \end{tabular}
  \label{table;HL_11minus}
\end{table*}
\renewcommand{\arraystretch}{1}


\section{Summary}\label{sec;summary}
In this paper, we analyzed the doubly heavy tetraquarks  
as a hadronic molecule of 
two open-heavy mesons. 
In the HQL, the heavy pseudoscalar 
and the heavy vector mesons are degenerate because of the HQS.  
Thus we took into account 
of possible $P^{(\ast)}P^{(\ast)}$ channel couplings. 
\\
\indent
As for $T_{cc}$ which has been reported by LHCb in 2022, we considered one meson exchange force 
where the cutoff parameter $\Lambda$ is determined to reproduce the experimental value of the $T_{cc}$ binding energy for $I(J^P)=0(1^+)$. 
However, in the case of the OPEP model, we were not able to find a bound state with the reasonable cutoff $\Lambda$. 
By adding the $\rho$, $\omega$, $\sigma$ exchanges (OBEP), we were able to obtain $\Lambda = 1069.8\,\mathrm{MeV}$ which reproduces the experimental value of $T_{cc}$. 
We also calculated the bound-state properties, the wave functions, mixing ratios and $\sqrt{\braket{r^2}}$, and it can be seen the channel ${[DD^{*}]}_{-}({}^3S_{1})$ is a dominant one. Next, we discussed $g_{s}$ 
dependence of the binding energy because the coupling constant $g_{s}$ for the $\sigma$ meson is uncertainly. 
We obtained that this dependence is large. However, by tuning the value of $\Lambda$ within a reasonable range, we found the set of $(g_s,\Lambda)$ which reproduces the experimental binding energy of $T_{cc}$.
Also, we studied bound states of $T_{cc}$ with 
given $I(J^{P})$ 
other than $0(1^{+})$. 
However we found no bound states of $T_{cc}$  
except for $0(1^{+})$.  
\\
\indent
As for $T_{bb}$ in the bottom sector, 
the $T_{bb}$ bound state with $0(1^{+})$ was found for $\Lambda\geq 1010\,\mathrm{MeV}$ even in the OPEP model. 
This enables us to expect that the $T_{bb}$ bound state with $0(1^{+})$ is likely to exist. 
Also using the OBEP, we calculated the binding energy of $T_{bb}$ with $0(1^{+})$ for $\Lambda=1069.8\,\mathrm{MeV}$, which reproduces the experimental value of $T_{cc}$. 
The binding energy of $T_{bb}$ is $46.0\,\mathrm{MeV}$ and we also obtained the wave functions, 
mixing ratios and $\sqrt{\braket{r^2}}$. 
Then, we found the ${[BB^{*}]}_{-}({}^3S_{1})$ channel is dominant and in addition the $B^{*}B^{*}({}^3S_{1})$ one is also important unlike $T_{cc}$. This reason can be understood because the mass difference of $B$ and $B^{*}$ is smaller than that of $D$ and $D^{*}$. 
This small mass difference 
leads to the deeply bound state in $T_{bb}$. Next, we calculated the bound states of $T_{bb}$ with given $I(J^{P})$ other than $0(1^{+})$ and got 
bound states. $g_{s}$ dependence of these bound states is different between $I=0$ and $I=1$.  
As $g_{s}$ increases, the binding energy decreases for $I=0$, but the binding energy increases for $I=1$. This difference is caused by the isospin factor $\vec{\tau}_{1}\cdot\vec{\tau}_{2}$. For $I=0$, the $\pi$ and $\rho$ exchange potentials contribute significantly in addition to 
the $\sigma$ exchange potential. Therefore, as $g_{s}$ increases, i.e., $\Lambda$ decreases, the binding energy becomes smaller. On the other hand, for $I=1$, since the $\sigma$ exchange potential is dominant, the binding energy is larger as $g_{s}$ increases. 
This enables us to study the $\sigma$ exchange potential in detail by searching $T_{bb}$ for $I=1$. \\
\indent
Finally, we considered the spin multiplets of the bound states in the HQL. We have reviewed the light-cloud basis and applied it to $T_{cc}$ with given $I(J^{P})$. 
In the HQL, we obtain many 
bound states for each quantum number, and were able to find that some pairs with different $J$ were degenerate.
In this analysis, we found that in the HQL 
the origin of $T_{cc}$ with $0(1^{+})$, which is reported by the LHCb, had the spin structure $S_{Q}=0, S_{q}=1$. Thus this state belongs to the HQS singlet, which has no HQS partner in the HQL. 
We were also able to see that the origin of the bound states of $T_{bb}$ with $0(0^{-})$, $1(0^{+})$ and $1(1^{+})$ have the spin 
structures $(S_{Q},S_{q},J_l)=(1,1,1)$, 
$(0,0,0)$ and 
$(1,1,1)$, respectively. In the HQL, every bound state of $T_{QQ}$ with $0(0^{-})$ is degenerate with a certain bound state of $T_{QQ}$ with $0(1^{-})$. 
In the bottom sector, however, the bound state of $T_{bb}$ with $0(1^{-})$ does not exist in our study, 
while the resonant state may exist. 
In the isovector channels, we obtained the $T_{bb}$ bound states with $1(0^{+})$ and $1(1^{+})$. 
The spin structure in the HQL showed that the origins of these bound states in the HQL were different. In fact, the state of $T_{bb}$ with $1(0^+)$ is a singlet state, while the state of $T_{bb}$ with $1(1^{+})$ is a triplet state. Therefore, the resonant states of $T_{bb}$ with $1(0^{+})$ and $1(2^+)$ which are partners of $1(1^+)$ may exist. \\
\indent
We only have considered the bound states in this study. As future works, we will investigate the resonant states of $T_{cc}$ and $T_{bb}$ and check the predictions noted above. 
We expect to find the bound states and the resonant states of the $P^{(\ast)}P^{(\ast)}$ hadronic molecules in future experiments. 

\section*{Acknowledgment}
This work is in part supported by Grants-in-Aid for Scientific Research under Grant Numbers JP20K14478 (Y.Y.).

\appendix
\section{Hamiltonian Matrix}\label{appendix;H}
In this section, we show the kinetic and potential matrices for given $I(J^P)$ in the hadronic-molecule basis~\cite{Ohkoda:2012hv}.
\subsection{kinetic energy matrix}\label{appendix;K}
The kinetic energy matrices for given $I(J^{P})$ are 
\begin{align*}
  K_{0(0^{-})} &= \diag\left({-\frac{1}{2\mu_{PP^{*}}}\triangle_{1}}\right),\\
  K_{0(1^{+})} &= \diag\left({-\frac{1}{2\mu_{PP^{*}}}\triangle_{0},-\frac{1}{2\mu_{PP^{*}}}\triangle_{2},-\frac{1}{2\mu_{P^*P^{*}}}\triangle_{0}+\Delta m_{PP^{*}},-\frac{1}{2\mu_{P^{*}P^{*}}}\triangle_{2}+\Delta m_{PP^{*}}}\right),\\
  K_{0(1^{-})} &= \operatorname{diag}\left(-\frac{1}{2\mu_{PP}}\triangle_{1},-\frac{1}{2\mu_{PP^{*}}}\triangle_{1}+\Delta m_{PP^{*}},-\frac{1}{2\mu_{P^{*}P^{*}}}\triangle_{1}+2\Delta m_{PP^{*}},\right.\\
  &\quad \left.-\frac{1}{2\mu_{P^{*}P^{*}}}\triangle_{1}+2m_{PP^{*}},-\frac{1}{2\mu_{P^{*}P^{*}}}\triangle_{3}+2m_{PP^{*}}\right),\\
  K_{1(0^{+})} &= \diag\left({-\frac{1}{2\mu_{PP}}\triangle_{0},-\frac{1}{2\mu_{P^{*}P^{*}}}\triangle_{0}+2\Delta m_{PP^{*}},-\frac{1}{2\mu_{P^{*}P^{*}}}\triangle_{2}+2\Delta m_{PP^{*}}}\right),\\
  K_{1(0^{-})} &= \diag\left({-\frac{1}{2\mu_{PP^{*}}}\triangle_{1},-\frac{1}{2\mu_{P^{*}P^{*}}}\triangle_{1}+\Delta m_{PP^{*}}}\right),\\
  K_{1(1^{+})} &= \diag\left({-\frac{1}{2\mu_{PP^{*}}}\triangle_{0},-\frac{1}{2\mu_{PP^{*}}}\triangle_{2},-\frac{1}{2\mu_{P^{*}P^{*}}}\triangle_{2}+\Delta m_{PP^{*}}}\right),\\
  K_{1(1^{-})} &= \diag\left({-\frac{1}{2\mu_{PP^{*}}}\triangle_{1},-\frac{1}{2\mu_{P^{*}P^{*}}}\triangle_{1}+\Delta m_{PP^{*}}}\right),
\end{align*}
where 
\begin{align*}
  \mu_{P^{(*)}P^{(*)}} &= \frac{m_{P^{(*)}}m_{P^{(*)}}}{m_{P^{(*)}} + m_{P^{(*)}}},\\
  \triangle_{l} &= \frac{d^2}{dr^2} - \frac{l(l+1)}{r^2},\\
  \Delta m_{PP^{*}} &= m_{P^{*}} - m_{P}.
\end{align*}
\subsection{Potential matrix in the hadronic-molecule basis}\label{appendix;V_HM}
The potential matrices in the hadronic-molecule basis are shown as follows:
\begin{itemize}
  \item $0(1^{+})$
  \begin{align*}
    V^{\mathrm{HM}}_{\pi,0(1^{+})} 
    &=\begin{pmatrix}
      -C_{\pi}&\!\!\sqrt{2}T_{\pi}&\!\!2C_{\pi}&\!\!\sqrt{2}T_{\pi}\\\vspace{1mm}
      \sqrt{2}T_{\pi}&\!\!-C_{\pi}-T_{\pi}&\!\!\sqrt{2}T_{\pi}&\!\!2C_{\pi}-T_{\pi}\\\vspace{1mm}
      2C_{\pi}&\!\!\sqrt{2}T_{\pi}&\!\!-C_{\pi}&\!\!\sqrt{2}T_{\pi}\\\vspace{1mm}
      \sqrt{2}T_{\pi}&\!\!2C_{\pi}-T_{\pi}&\!\!\sqrt{2}T_{\pi}&\!\!-C_{\pi}-T_{\pi}
    \end{pmatrix},\\
    V^{\mathrm{HM}}_{v,0(1^{+})}
    &=\begin{pmatrix}
      C^{\prime}_{v}-2C_{v}&\!\!-\sqrt{2}T_{v}&\!\!4C_{v}&\!\!-\sqrt{2}T_{v}\\\vspace{1mm}
      -\sqrt{2}T_{v}&\!\!C^{\prime}_{v}-2C_{v}+T_{v}&\!\!-\sqrt{2}T_{v}&\!\!4C_{v}+T_{v}\\\vspace{1mm}
      4C_{v}&\!\!-\sqrt{2}T_{v}&\!\!C^{\prime}_{v}-2C_{v}&\!\!-\sqrt{2}T_{v}\\\vspace{1mm}
      -\sqrt{2}T_{v}&\!\!4C_{v}+T_{v}&\!\!-\sqrt{2}T_{v}&\!\!C^{\prime}_{v}-2C_{v}+T_{v}
    \end{pmatrix},\\
    V^{\HM}_{\sigma,0(1^{+})} 
    &= \begin{pmatrix}
      C_{\sigma}&0&0&0\\
      0&C_{\sigma}&0&0\\
      0&0&C_{\sigma}&0\\
      0&0&0&C_{\sigma}
    \end{pmatrix},
  \end{align*} 
  \item $0(0^{-})$
  \begin{align*}
    &V^{\mathrm{HM}}_{\pi,0(0^{-})} 
    = (C_{\pi}+2T_{\pi}),\\
    &V^{\mathrm{HM}}_{v,0(0^{-})} 
    = (C^{\prime}_{v}+2C_{v}-2T_{v}),\\
    &V^{\mathrm{HM}}_{\sigma,0(0^{-})}
    = (C_{\sigma}),
  \end{align*}  
  \item $0(1^{-})$
  \begin{align*}
    V^{\mathrm{HM}}_{\pi,0(1^{-})} &= \begin{pmatrix}
      0&\!\!0&\!\!-\sqrt{3}C_{\pi}&\!\!-2\sqrt{\frac{3}{5}}T_{\pi}&3\sqrt{\frac{2}{5}}T_{\pi}\\\vspace{1mm}
      0&\!\!C_{\pi}-T_{\pi}&\!\!0&\!\!-3\sqrt{\frac{3}{5}}T_{\pi}&\!\!-3\sqrt{\frac{2}{5}}T_{\pi}\\\vspace{1mm}
      -\sqrt{3}C_{\pi}&\!\!0&\!\!-2C_{\pi}&\!\!\frac{2}{\sqrt{5}}T_{\pi}&\!\!-\sqrt{\frac{6}{5}}T_{\pi}\\\vspace{1mm}
      -2\sqrt{\frac{3}{5}}T_{\pi}&\!\!-3\sqrt{\frac{3}{5}}T_{\pi}&\!\!\frac{2}{\sqrt{5}}T_{\pi}&\!\!C_{\pi}-\frac{7}{5}T_{\pi}&\!\!\frac{\sqrt{6}}{5}T_{\pi}\\\vspace{1mm}
      3\sqrt{\frac{2}{5}}T_{\pi}&\!\!-3\sqrt{\frac{2}{5}}T_{\pi}&\!\!-\sqrt{\frac{6}{5}}T_{\pi}&\!\!\frac{\sqrt{6}}{5}T_{\pi}&\!\!C_{\pi}-\frac{8}{5}T_{\pi}
    \end{pmatrix},\\
    V^{\mathrm{HM}}_{v,0(1^{-})} &= \begin{pmatrix}
      C^{\prime}_{v}&\!\!0&\!\!-2\sqrt{3}C_{v}&\!\!2\sqrt{\frac{3}{5}}T_{v}&-3\sqrt{\frac{2}{5}}T_{v}\\\vspace{1mm}
      0&\!\!C^{\prime}_{v}+2C_{v}+T_{v}&\!\!0&\!\!3\sqrt{\frac{3}{5}}T_{v}&\!\!3\sqrt{\frac{2}{5}}T_{v}\\\vspace{1mm}
      -2\sqrt{3}C_{v}&\!\!0&\!\!C^{\prime}_{v}-4C_{v}&\!\!-\frac{2}{\sqrt{5}}T_{v}&\!\!\sqrt{\frac{6}{5}}T_{v}\\\vspace{1mm}
      2\sqrt{\frac{3}{5}}T_{v}&\!\!3\sqrt{\frac{3}{5}}T_{v}&\!\!-\frac{2}{\sqrt{5}}T_{v}&\!\!C^{\prime}_{v}+2C_{v}+\frac{7}{5}T_{v}&\!\!-\frac{\sqrt{6}}{5}T_{v}\\\vspace{1mm}
      -3\sqrt{\frac{2}{5}}T_{v}&\!\!3\sqrt{\frac{2}{5}}T_{v}&\!\!\sqrt{\frac{6}{5}}T_{v}&\!\!-\frac{\sqrt{6}}{5}T_{v}&\!\!C^{\prime}_{v}+2C_{v}+\frac{8}{5}T_{v}
    \end{pmatrix},\\
    V^{\HM}_{\sigma,0(1^{-})} &= \begin{pmatrix}
      C_{\sigma}&0&0&0&0\\
      0&C_{\sigma}&0&0&0\\
      0&0&C_{\sigma}&0&0\\
      0&0&0&C_{\sigma}&0\\
      0&0&0&0&C_{\sigma}
    \end{pmatrix},
  \end{align*}  
  \item $1(0^{+})$
  \begin{align*}
    V^{\mathrm{HM}}_{\pi,1(0^{+})} &= \begin{pmatrix}
      0&\!\!-\sqrt{3}C_{\pi}&\!\!\sqrt{6}T_{\pi}\\\vspace{1mm}
      -\sqrt{3}C_{\pi}&\!\!-2C_{\pi}&\!\!-\sqrt{2}T_{\pi}\\\vspace{1mm}
      \sqrt{6}T_{\pi}&\!\!-\sqrt{2}T_{\pi}&\!\!C_{\pi}-2T_{\pi}
    \end{pmatrix},\\
    V^{\mathrm{HM}}_{v,1(0^{+})} &= \begin{pmatrix}
      C^{\prime}_{v}&\!\!-2\sqrt{3}C_{v}&\!\!-\sqrt{6}T_{v}\\\vspace{1mm}
      -2\sqrt{3}C_{v}&\!\!C^{\prime}_{v}-4C_{v}&\!\!\sqrt{2}T_{v}\\\vspace{1mm}
      -\sqrt{6}T_{v}&\!\!\sqrt{2}T_{v}&\!\!C^{\prime}_{v}+2C_{v}+2T_{v}
    \end{pmatrix},\\
    V^{\HM}_{\sigma,1(0^{+})} &= \begin{pmatrix}
      C_{\sigma}&0&0\\
      0&C_{\sigma}&0\\
      0&0&C_{\sigma}
    \end{pmatrix},
  \end{align*}  
  \item $1(0^{-})$
  \begin{align*}
    V^{\mathrm{HM}}_{\pi,1(0^{-})} &= \begin{pmatrix}
      -C_{\pi}-2T_{\pi}&\!\!2C_{\pi}-2T_{\pi}\\\vspace{1mm}
      2C_{\pi}-2T_{\pi}&\!\!-C_{\pi}-2T_{\pi}
    \end{pmatrix},\\
    V^{\mathrm{HM}}_{\pi,1(0^{-})} &= \begin{pmatrix}
      C^{\prime}_{v}-2C_{v}+2T_{v}&\!\!4C_{v}+2T_{v}\\\vspace{1mm}
      4C_{v}+2T_{v}&\!\!C^{\prime}_{v}-2C_{v}+2T_{v}
    \end{pmatrix},\\
    V^{\HM}_{\sigma,1(0^{-})} &= \begin{pmatrix}
      C_{\sigma}&0\\
      0&C_{\sigma}
    \end{pmatrix},
  \end{align*}  
  \item $1(1^{+})$
  \begin{align*}
      V^{\mathrm{HM}}_{\pi,1(1^{+})} &= \begin{pmatrix}
        C_{\pi}&\!\!-\sqrt{2}T_{\pi}&\!\!-\sqrt{6}T_{\pi}\\\vspace{1mm}
        -\sqrt{2}T_{\pi}&\!\!C_{\pi}+T_{\pi}&\!\!-\sqrt{3}T_{\pi}\\\vspace{1mm}
        -\sqrt{6}T_{\pi}&\!\!-\sqrt{3}T_{\pi}&\!\!C_{\pi}-T_{\pi}
      \end{pmatrix},\\
      V^{\mathrm{HM}}_{v,1(1^{+})} &= \begin{pmatrix}
        C^{\prime}_{v}+2C_{v}&\!\!\sqrt{2}T_{v}&\!\!\sqrt{6}T_{v}\\\vspace{1mm}
        \sqrt{2}T_v&\!\!C^{\prime}_{v}+2C_{v}-T_{v}&\!\!\sqrt{3}T_{v}\\\vspace{1mm}
        \sqrt{6}T_{v}&\!\!\sqrt{3}T_{v}&\!\!C^{\prime}_{v}+2C_{v}+T_{v}
      \end{pmatrix},\\
      V^{\HM}_{\sigma,1(1^{+})} &= \begin{pmatrix}
        C_{\sigma}&0&0\\
        0&C_{\sigma}&0\\
        0&0&C_{\sigma}
      \end{pmatrix},
  \end{align*}  
  \item $1(1^{-})$
  \begin{align*}
    V^{\mathrm{HM}}_{\pi,1(1^{-})} &= \begin{pmatrix}
      -C_{\pi}+T_{\pi}&\!\!2C_{\pi}+T_{\pi}\\\vspace{1mm}
      2C_{\pi}+T_{\pi}&\!\!-C_{\pi}+T_{\pi}
    \end{pmatrix},\\
    V^{\mathrm{HM}}_{\pi,1(1^{-})} &= \begin{pmatrix}
      C^{\prime}_{v}-2C_{v}-T_{v}&\!\!4C_{v}-T_{v}\\\vspace{1mm}
      4C_{v}-T_{v}&\!\!C^{\prime}_{v}-2C_{v}-T_{v}
    \end{pmatrix},\\
    V^{\HM}_{\sigma,1(1^{-})} &= \begin{pmatrix}
      C_{\sigma}&0\\
      0&C_{\sigma}
    \end{pmatrix},
  \end{align*}
  \begin{align*}
    C_{\pi} &= \frac{1}{3}{\left(\frac{g}{2f_{\pi}}\right)}^2 C(r;m_{\pi})\vec{\tau}_{1}\cdot\vec{\tau}_{2},\\
    T_{\pi} &= \frac{1}{3}{\left(\frac{g}{2f_{\pi}}\right)}^2 T(r;m_{\pi})\vec{\tau}_{1}\cdot\vec{\tau}_{2},\\
    C^{\prime}_{v} &= {\left(\frac{\beta g_{V}}{2m_{v}}\right)}^2 C(r;m_{v})\vec{\tau}_{1}\cdot\vec{\tau}_{2},\\
    C_{v} &= \frac{1}{3}{(\lambda g_{V})}^2 C(r;m_{v})\vec{\tau}_{1}\cdot\vec{\tau}_{2},\\
    T_{v} &= \frac{1}{3}{(\lambda g_{V})}^2 T(r;m_{v})\vec{\tau}_{1}\cdot\vec{\tau}_{2},\\
    C_{\sigma} &= -{\left(\frac{g}{m_{\sigma}}\right)}^2 C(r;m_{\sigma}).
  \end{align*}
\end{itemize}

\section{Light-Cloud Basis}\label{appendix;LCB}
The possible channels, the light-cloud bases and the potential matrices in the light-cloud basis are shown as follows:
\begin{itemize}
  \item $0(1^{+})$
  \begin{align*}
    {\psi^{\mathrm{HM}}_{0(1^{+})}} &= 
    \begin{pmatrix}
      \ket{{[PP^{*}]}_{-}({}^3S_{1})}\vspace{1mm}\\\ket{{[PP^{*}]}_{-}({}^3D_{1})}\vspace{1mm}\\
      \ket{P^{*}P^{*}({}^3S_{1})}\vspace{1mm}\\\ket{P^{*}P^{*}({}^3D_{1})}
    \end{pmatrix},\\
    \psi^{\mathrm{LC}}_{0(1^{+})} 
    &= U^{-1}_{0(1^{+})} \psi^{\mathrm{HM}}_{0(1^{+})}\notag \\
    &= 
    \begin{pmatrix}
      {\Ket{{\Big[{\big[QQ\big]}_{1}\ {\big[S\ {[\bar{q}\bar{q}]}_{0}\big]}_{0}\Big]}_{1}}}\vspace{1mm}\\
      {\Ket{{\Big[{\big[QQ\big]}_{0}\ {\big[S\ {[\bar{q}\bar{q}]}_{1}\big]}_{1}\Big]}_{1}}}\vspace{1mm}\\
      {\Ket{{\Big[{\big[QQ\big]}_{0}\ {\big[D\ {[\bar{q}\bar{q}]}_{1}\big]}_{1}\Big]}_{1}}}\vspace{1mm}\\
      {\Ket{{\Big[{\big[QQ\big]}_{1}\ {\big[D\ {[\bar{q}\bar{q}]}_{0}\big]}_{2}\Big]}_{1}}}\vspace{1mm}
    \end{pmatrix},\\
    U_{0(1^{+})} &= \begin{pmatrix}
      -\frac{1}{\sqrt{2}}&\frac{1}{\sqrt{2}}&0&0\vspace{1mm}\\
      0&0&\frac{1}{\sqrt{2}}&-\frac{1}{\sqrt{2}}\vspace{1mm}\\
      \frac{1}{\sqrt{2}}&\frac{1}{\sqrt{2}}&0&0\vspace{1mm}\\
      0&0&\frac{1}{\sqrt{2}}&\frac{1}{\sqrt{2}}
    \end{pmatrix},\\
    V^{\mathrm{LC}}_{\pi,0(1^{+})} &= U^{-1}_{0(1^{+})}V^{\mathrm{HM}}_{\pi,0(1^{+})}U_{0(1^{+})}\notag\\
    &=\left(\begin{array}{c|cc|c}
      -3C_{\pi}&0&0&0\\
      \hline
      0&C_{\pi}&2\sqrt{2}T_{\pi}&0\\
      0&2\sqrt{2}T_{\pi}&C_{\pi}-2T_{\pi}&0\\
      \hline
      0&0&0&-3C_{\pi}
    \end{array}\right),\\
    V^{\mathrm{LC}}_{v,0(1^{+})} &= U^{-1}_{0(1^{+})}V^{\mathrm{HM}}_{v,0(1^{+})}U_{0(1^{+})}\notag\\
    &=\left(
      \begin{array}{c|cc|c}
        C^{\prime}_{v}-6C_{v}&0&0&0\\\hline
        0&C^{\prime}_{v}+2C_{v}&-2\sqrt{2}T_{v}&0\\
        0&-2\sqrt{2}T_{v}&C^{\prime}_{v}+2C_{v}+2T_{v}&0\\\hline
        0&0&0&C^{\prime}_{v}-6C_{v}
      \end{array}
    \right),\\
    V^{\mathrm{LC}}_{\sigma,0(1^{+})} &= U^{-1}_{0(1^{+})}V^{\mathrm{HM}}_{\sigma,0(1^{+})}U_{0(1^{+})}\notag\\
    &=\left(
      \begin{array}{c|cc|c}
        C_{\sigma}&0&0&0\\\hline
        0&C_{\sigma}&0&0\\
        0&0&C_{\sigma}&0\\\hline
        0&0&0&C_{\sigma}
      \end{array}
    \right),
  \end{align*}
  \item $0(0^{-})$
  \begin{align*}
    &\psi^{\mathrm{HM}}_{0(0^{-})} = \begin{pmatrix}
      \ket{{[PP^{*}]}_{+}({}^3P_{0})}
    \end{pmatrix},\\
    &\psi^{\mathrm{LC}}_{0(0^{-})} = \begin{pmatrix}
      -\Ket{{\Big[{\big[QQ\big]}_{1}\ {\big[P\ {[\bar{q}\bar{q}]}_{1}\big]}_{1}\Big]}_{0}}
    \end{pmatrix},\\
    &V^{\mathrm{HM}}_{\pi,0(0^{-})} 
    = (C_{\pi}+2T_{\pi}),\\
    &V^{\mathrm{HM}}_{v,0(0^{-})} 
    = (C^{\prime}_{v}+2C_{v}-2T_{v}),\\
    &V^{\HM}_{\sigma,0(0^{-})}
    = C_{\sigma},
  \end{align*}
  \item $0(1^{-})$
  \begin{align*}
    \psi^{\mathrm{HM}}_{0(1^{-})} &= \begin{pmatrix}
      \ket{PP({}^1P_{1})}\vspace{1mm}\\
      \ket{{[PP^{*}]}_{+}({}^3P_{1})}\vspace{1mm}\\
      \ket{P^{*}P^{*}({}^1P_{1})}\vspace{1mm}\\
      \ket{P^{*}P^{*}({}^5P_{1})}\vspace{1mm}\\
      \ket{P^{*}P^{*}({}^5F_{1})}
    \end{pmatrix},\\
    \psi^{\mathrm{LC}}_{0(1^{-})} &= U^{-1}_{0(1^{-})}\psi^{\mathrm{HM}}_{0(1^{-})}\\
    &= \begin{pmatrix}
      \Ket{{\Big[{\big[QQ\big]}_{0}\ {\big[P\ {[\bar{q}\bar{q}]}_{0}\big]}_{1}\Big]}_{1}}\vspace{1mm}\\
      \Ket{{\Big[{\big[QQ\big]}_{1}\ {\big[P\ {[\bar{q}\bar{q}]}_{1}\big]}_{0}\Big]}_{1}}\vspace{1mm}\\
      \Ket{{\Big[{\big[QQ\big]}_{1}\ {\big[P\ {[\bar{q}\bar{q}]}_{1}\big]}_{1}\Big]}_{1}}\vspace{1mm}\\
      \Ket{{\Big[{\big[QQ\big]}_{1}\ {\big[P\ {[\bar{q}\bar{q}]}_{1}\big]}_{2}\Big]}_{1}}\vspace{1mm}\\
      \Ket{{\Big[{\big[QQ\big]}_{1}\ {\big[F\ {[\bar{q}\bar{q}]}_{1}\big]}_{2}\Big]}_{1}}
    \end{pmatrix},\\
    U_{0(1^{-})} &= \begin{pmatrix}
      \frac{1}{2}&\frac{\sqrt{3}}{6}&\frac{1}{2}&\frac{\sqrt{15}}{6}&0\vspace{1mm}\\
      0&\frac{\sqrt{3}}{3}&\frac{1}{2}&-\frac{\sqrt{15}}{6}&0\vspace{1mm}\\
      \frac{\sqrt{3}}{2}&-\frac{1}{6}&-\frac{\sqrt{3}}{6}&-\frac{\sqrt{5}}{6}&0\vspace{1mm}\\
      0&\frac{\sqrt{5}}{3}&-\frac{\sqrt{15}}{6}&\frac{1}{6}&0\vspace{1mm}\\
      0&0&0&0&1
    \end{pmatrix},\\
    V^{\mathrm{LC}}_{\pi,0(1^{-})} &= \left(\begin{array}{c|c|c|cc}
      -3C_{\pi}&0&0&0&0\\
      \hline
      0&C_{\pi}-4T_{\pi}&0&0&0\\
      \hline
      0&0&C_{\pi}+2T_{\pi}&0&0\\
      \hline
      0&0&0&C_{\pi}-\frac{2}{5}T_{\pi}&\frac{6\sqrt{6}}{5}T_{\pi}\\
      0&0&0&\frac{6\sqrt{6}}{5}T_{\pi}&C_{\pi}-\frac{8}{5}T_{\pi}
    \end{array}\right),\\
    V^{\mathrm{LC}}_{v,0(1^{-})} &= \left(
      \begin{array}{c|c|c|cc}
        C^{\prime}_{v}-6C_{v}&0&0&0&0\\\hline
        0&C^{\prime}_{v}+2C_{v}+4T_{v}&0&0&0\\\hline
        0&0&C^{\prime}_{v}+2C_{v}-2T_{v}&0&0\\\hline
        0&0&0&C^{\prime}_{v}+2C_{v}+\frac{2}{5}T_{v}&-\frac{6\sqrt{6}}{5}T_{v}\\
        0&0&0&-\frac{6\sqrt{6}}{5}T_{v}&\!\!C^{\prime}_{v}+2C_{v}+\frac{8}{5}T_{v}
      \end{array}
    \right),\\
    V^{\LC}_{\sigma,0(1^{-})} &= \left(
      \begin{array}{c|c|c|cc}
        C_{\sigma}&0&0&0&0\\\hline
        0&C_{\sigma}&0&0&0\\\hline
        0&0&C_{\sigma}&0&0\\\hline
        0&0&0&C_{\sigma}&0\\
        0&0&0&0&C_{\sigma}
      \end{array}
    \right),
  \end{align*}
  \item $1(0^{+})$
  \begin{align*}
    \psi^{\mathrm{HM}}_{1(0^{+})} &= \begin{pmatrix}
      \ket{PP({}^1S_{0})}\vspace{1mm}\\
      \ket{P^{*}P^{*}({}^1S_{0})}\vspace{1mm}\\
      \ket{P^{*}P^{*}({}^5D_{0})}\vspace{1mm}\\
    \end{pmatrix},\\
    \psi^{\mathrm{LC}}_{1(0^{+})} &= U^{-1}_{1(0^{+})}\psi^{\mathrm{HM}}_{1(0^{+})}\notag\\
    &=\begin{pmatrix}
      \Ket{{\Big[{\big[QQ\big]}_{0}\ {\big[S\ {[\bar{q}\bar{q}]}_{0}\big]}_{0}\Big]}_{0}}\vspace{1mm}\\
      \Ket{{\Big[{\big[QQ\big]}_{1}\ {\big[S\ {[\bar{q}\bar{q}]}_{1}\big]}_{1}\Big]}_{0}}\vspace{1mm}\\
      \Ket{{\Big[{\big[QQ\big]}_{1}\ {\big[D\ {[\bar{q}\bar{q}]}_{1}\big]}_{1}\Big]}_{0}}
    \end{pmatrix},\\
    U_{1(0^{+})} &= 
      \begin{pmatrix}
        \frac{1}{2}&\frac{\sqrt{3}}{2}&0\vspace{1mm}\\
        \frac{\sqrt{3}}{2}&-\frac{1}{2}&0\vspace{1mm}\\
        0&0&1
      \end{pmatrix},\\
    V^{\mathrm{LC}}_{\pi,1(0^{+})} &= \left(\begin{array}{c|cc}
      -3C_{\pi}&0&0\\\hline
      0&C_{\pi}&2\sqrt{2}T_{\pi}\\
      0&2\sqrt{2}T_{\pi}&C_{\pi}-2T_{\pi}
    \end{array}\right),\\
    V^{\mathrm{LC}}_{v,1(0^{+})} &= \left(
      \begin{array}{c|cc}
        C^{\prime}_{v}-6C_{v}&0&0\\\hline
        0&C^{\prime}_{v}+2C_{v}&-2\sqrt{2}T_{v}\\
        0&-2\sqrt{2}T_{v}&C^{\prime}_{v}+2C_{v}+2T_{v}
      \end{array}
    \right),\\
    V^{\LC}_{\sigma,1(0^{+})} &= \left(
      \begin{array}{c|cc}
        C_{\sigma}&0&0\\\hline
        0&C_{\sigma}&0\\
        0&0&C_{\sigma}
      \end{array}
    \right),
  \end{align*}
  \item $1(0^{-})$
  \begin{align*}
    \psi^{\mathrm{HM}}_{1(0^{-})} &= \begin{pmatrix}
      \ket{{[PP^{*}]}_{-}({}^3P_{0})}\\
      \ket{P^{*}P^{*}({}^3P_{0})}
    \end{pmatrix},\\
    \psi^{\mathrm{LC}}_{1(0^{-})} &= U^{-1}_{1(0^{-})}\psi^{\mathrm{HM}}_{1(0^{-})}\notag\\
    &=\begin{pmatrix}
      \Ket{{\Big[{\big[QQ\big]}_{0}\ {\big[P\ {[\bar{q}\bar{q}]}_{1}\big]}_{0}\Big]}_{0}}\vspace{1mm}\\
      \Ket{{\Big[{\big[QQ\big]}_{1}\ {\big[P\ {[\bar{q}\bar{q}]}_{0}\big]}_{1}\Big]}_{0}}
    \end{pmatrix},\\
    U_{1(0^{-})} &= 
    \begin{pmatrix}
      \frac{1}{\sqrt{2}}&-\frac{1}{\sqrt{2}}\vspace{1mm}\\
      \frac{1}{\sqrt{2}}&\frac{1}{\sqrt{2}}
    \end{pmatrix},\\
    V^{\mathrm{LC}}_{\pi,1(0^{-})} &= \left(
      \begin{array}{c|c}
        C_{\pi}-4T_{\pi}&0\\\hline
        0&-3C_{\pi}
      \end{array}
    \right),\\
    V^{\mathrm{LC}}_{\pi,1(0^{-})} &= \left(
      \begin{array}{c|c}
        C^{\prime}_{v}+2C_{v}+4T_{v}&0\\\hline
        0&C^{\prime}_{v}-6C_{v}
      \end{array}
    \right),\\
    V^{\LC}_{\sigma,1(0^{-})} &= \left(
      \begin{array}{c|c}
        C_{\sigma}&0\\\hline
        0&C_{\sigma}
      \end{array}
    \right),
  \end{align*}
  \item $1(1^{+})$
  \begin{align*}
    \psi^{\mathrm{HM}}_{1(1^{+})} &= \begin{pmatrix}
      \ket{{[PP^{*}]}_{+}({}^3S_{1})}\vspace{1mm}\\
      \ket{{[PP^{*}]}_{+}({}^3D_{1})}\vspace{1mm}\\
      \ket{P^{*}P^{*}({}^5D_{1})}\vspace{1mm}\\
    \end{pmatrix},\\
    \psi^{\mathrm{LC}}_{1(1^{+})} &= U^{-1}_{1(1^{+})}\psi^{\mathrm{HM}}_{1(0^{+})}\notag\\
    &=\begin{pmatrix}
      \Ket{{\Big[{\big[QQ\big]}_{1}\ {\big[S\ {[\bar{q}\bar{q}]}_{1}\big]}_{1}\Big]}_{1}}\vspace{1mm}\\
      \Ket{{\Big[{\big[QQ\big]}_{1}\ {\big[D\ {[\bar{q}\bar{q}]}_{1}\big]}_{1}\Big]}_{1}}\vspace{1mm}\\
      \Ket{{\Big[{\big[QQ\big]}_{1}\ {\big[D\ {[\bar{q}\bar{q}]}_{1}\big]}_{2}\Big]}_{1}}
    \end{pmatrix},\\
    U_{1(1^{+})} &= 
      \begin{pmatrix}
        1&0&0\vspace{1mm}\\
        0&-\frac{1}{2}&-\frac{\sqrt{3}}{2}\vspace{1mm}\\
        0&-\frac{\sqrt{3}}{2}&\frac{1}{2}
      \end{pmatrix},\\
    V^{\mathrm{LC}}_{\pi,1(1^{+})} &= \left(\begin{array}{cc|c}
      C_{\pi}&2\sqrt{2}T_{\pi}&0\\
      2\sqrt{2}T_{\pi}&C_{\pi}-2T_{\pi}&0\\\hline
      0&0&C_{\pi}+2T_{\pi}
    \end{array}\right),\\
    V^{\mathrm{LC}}_{v,1(1^{+})} &= \left(
      \begin{array}{cc|c}
        C^{\prime}_{v}+2C_{v}&-2\sqrt{2}T_{v}&0\\
        -2\sqrt{2}T_{v}&C^{\prime}_{v}+2C_{v}+2T_{v}&0\\\hline
        0&0&C^{\prime}_{v}+2C_{v}-2T_{v}
      \end{array}
    \right),\\
    V^{\LC}_{\sigma,1(1^{+})} &= \left(
      \begin{array}{cc|c}
        C_{\sigma}&0&0\\
        0&C_{\sigma}&0\\\hline
        0&0&C_{\sigma}
      \end{array}
    \right),
  \end{align*}
  \item $1(1^{-})$
  \begin{align*}
    \psi^{\mathrm{HM}}_{1(1^{-})} &= \begin{pmatrix}
      \ket{{[PP^{*}]}_{-}({}^3P_{1})}\\
      \ket{P^{*}P^{*}({}^3P_{1})}
    \end{pmatrix},\\
    \psi^{\mathrm{LC}}_{1(1^{-})} &= U^{-1}_{1(1^{-})}\psi^{\mathrm{HM}}_{1(1^{-})}\notag\\
    &=\begin{pmatrix}
      \Ket{{\Big[{\big[QQ\big]}_{0}\ {\big[P\ {[\bar{q}\bar{q}]}_{1}\big]}_{1}\Big]}_{1}}\vspace{1mm}\\
      \Ket{{\Big[{\big[QQ\big]}_{1}\ {\big[P\ {[\bar{q}\bar{q}]}_{0}\big]}_{1}\Big]}_{1}}
    \end{pmatrix},\\
    U_{1(1^{-})} &= 
    \begin{pmatrix}
      \frac{1}{\sqrt{2}}&\frac{1}{\sqrt{2}}\vspace{1mm}\\
      \frac{1}{\sqrt{2}}&-\frac{1}{\sqrt{2}}
    \end{pmatrix},\\
    V^{\mathrm{LC}}_{\pi,1(1^{-})} &= \left(
      \begin{array}{c|c}
        C_{\pi}+2T_{\pi}&0\\\hline
        0&-3C_{\pi}
      \end{array}
    \right),\\
    V^{\mathrm{LC}}_{\pi,1(1^{-})} &= \left(
      \begin{array}{c|c}
        C^{\prime}_{v}+2C_{v}-2T_{v}&0\\\hline
        0&C^{\prime}_{v}-6C_{v}
      \end{array}
    \right),\\
    V^{\LC}_{\sigma,1(1^{-})} &= \left(
      \begin{array}{c|c}
        C_{\sigma}&0\\\hline
        0&C_{\sigma}
      \end{array}
    \right).
  \end{align*}
\end{itemize}
\nocite{*}
\bibliography{reference}
\end{document}